\documentclass[11pt]{article}
\usepackage{amssymb}
\usepackage{amsmath}
\usepackage{amscd}
\usepackage{amsthm}

\numberwithin{equation}{section}

\newtheorem{prop}{Proposition}

\normalsize
\begin{document}
\bibliographystyle{plain}

\vspace{2cm}
\title{\bfseries Remarks on the thermodynamics and the vacuum energy of a quantum Maxwell gas on compact and
closed manifolds}

\author{Gerald Kelnhofer \\
Faculty of Physics \\
University of Vienna \\
Boltzmanngasse 5, A-1090 Vienna, \\ Austria}
\date{}
\maketitle
\begin{abstract}
The quantum Maxwell theory at finite temperature at equilibrium is studied on compact and closed manifolds in both the functional integral- and
Hamiltonian formalism. The aim is to shed some light onto the interrelation between the topology of the spatial background and the
thermodynamic properties of the system. The quantization is not unique and gives rise to inequivalent quantum theories which are classified by
$\theta$-vacua. Based on explicit parametrizations of the gauge orbit space in the functional integral approach and of the physical phase space
in the canonical quantization scheme, the Gribov problem is resolved and the equivalence of both quantization schemes is elucidated. Using
zeta-function regularization the free energy is determined and the effect of the topology of the spatial manifold on the vacuum energy and on
the thermal gauge field excitations is clarified. The general results are then applied to a quantum Maxwell gas on a $n$-dimensional torus
providing explicit formulae for the main thermodynamic functions in the low- and high temperature
regimes, respectively.
Report-No.: UWThPh-2012-3
\end{abstract}




\section{Introduction}

Quantum field theories at finite temperature have been intensively studied during the last years \cite{Kapusta,Bellac}. Prominent examples
which stimulate this interest are the study of matter formation in the early stage of the universe, the description of the quark gluon plasma
in the context of the AdS/CFT correspondence \cite{GKP} and the current intense and partly controversial discussion concerning the Casimir
effect \cite{Casimir} at finite temperature (for a review see \cite{PMG, MT, K-A-Milton, BMM, BKMM, BD}). From a general perspective, the
\textit{thermal Casimir effect} can be regarded as a deviation of the vacuum energy and the energy of thermal excitations of a quantum field
caused by the presence of external constraints. These constraints may be imposed either by real material boundaries or by topologically
non-trivial manifolds on which the quantum fields reside. In any case, the modes of the quantum fields are correspondingly restricted,
affecting both the vacuum energy and the thermodynamic functions of the system.\par

In particular, it is the relation between the topology of the spatial background and the thermodynamic properties of gauge fields which in our
opinion deserves closer attention and is the main motivation for the present paper. Our purpose is to study quantum Maxwell theory at finite
temperature at equilibrium on a $n$-dimensional compact, closed and connected manifold $X$, which represents the spatial background. We perform
the analysis in both the functional integral and Hamiltonian approach and derive the expression for the regularized free energy.\par

Let us now motivate our intention of the present paper in more detail: Quantum fields at finite temperature on general manifolds with and
without a boundary have been considered for many years \cite{DK1, D, Amb-Wolf, BCVZ, ZK1, ZK2, Spraefico, BMO, ET, D-2002, DX, edery, Lim-Teo,
GKM, SZ}. (Here we have selected a list of references which are the most relevant ones in the context we are interested in). Most of the
research is in fact dedicated to the study of scalar fields. It is often argued, however, that the thermodynamic functions for a photon gas
could be obtained directly from those for a massless scalar gas just by multiplication with the number of independent polarization states. We
will show explicitly that this statement is not true in general. This requires a critical review of the concept which underlies the description
of finite temperature gauge fields.\par

In the functional integral formulation the thermal partition function of a bosonic systems at finite temperature $T$ is obtained by integrating
the classical action over all fields periodic in the (imaginary) time coordinate with period $\beta :=\frac{1}{k_BT}$, where $k_B$ is the
Boltzmann constant \cite{Kapusta,Bellac}. Geometrically this corresponds to a quantum field theory of bosons on the product space
$\mathbb{T}_{\beta}^1\times X$, where $\mathbb{T}_{\beta}^1$ denotes the circle ($1$-torus) with circumference $\beta$.\par

Evidently, the partition function plays a major role in the finite temperature context, since it serves as basic quantity from which the
thermodynamic functions are derived. Hence particular care has to be taken in order to obtain the correct measure for the functional integral
representation of the partition function. This issue has been explicitly discussed in the seminal papers \cite{bernard,hata} for quantum
Maxwell theory in the covariant gauge in the non-compact but topologically trivial case $X=\mathbb{R}^n$. Following the Faddeev-Popov approach,
the corresponding Faddeev-Popov determinant which appears as a field independent but temperature dependent multiplicative factor after gauge
fixing must be retained in the partition function. Compared to the zero-temperature case, this factor is necessary to compensate the
contributions caused by the unphysical degrees of freedom.\par

But what happens if the thermal gauge theory suffers from Gribov ambiguities, which prevent the existence of a global and unique solution of
the gauge-fixing condition, and thus the existence of any globally defined measure on the space of thermal gauge fields?

It is a general result that abelian gauge theories suffer from the Gribov problem whenever the gauge fields reside on a non-simply connected,
compact manifold \cite{killingback,kelnhofer}. However, this is precisely the situation one encounters in the finite temperature context since
the base manifold is the product space $\mathbb{T}_{\beta}^1\times X$. Thus even for topologically trivial spatial manifolds $X$, this problem
does exist. This fact had been recognized long time ago \cite{Parthasarathy,Parthasarathy2} in the case of quantum Maxwell theory at finite
temperature on a $3$-sphere ($X=\mathbb{S}^3$) by solving the corresponding gauge-fixing condition. The main result was that the Faddeev-Popov
operator (i.e. the zeroth order Laplace operator on $\mathbb{T}_{\beta}^1\times \mathbb{S}^3$) possesses temperature dependent zero-modes.
Based on the results of Ref. \cite{bernard} it was then argued, but not proved, that the Faddeev-Popov formula for the partition function is
valid as long as the zero modes are ruled out from the domain of the Faddeev-Popov operator. Although this argument is valid for
$X=\mathbb{S}^3$, we will show, however, that restriction to the non-zero modes alone is not sufficient to give the correct partition function
in the general case.\par

In this paper the construction of the partition function and thus the Gribov problem will be tackled in a different way. As a by-product the
relation between the occurrence of the zero modes and the Gribov ambiguities is elucidated, too. Thereby we will focus on the following four
objectives:\par

\begin{itemize}
    \item Resolve the Gribov problem and construct a reasonable functional integral representation of the free energy including a finite expression
    for the vacuum energy using zeta function regularization.
    \item Analyze the relation between the functional integral quantization and the Hamiltonian (canonical) quantization at finite temperature.
    \item Study the impact of the topology and geometry of the spatial manifold $X$ on the thermodynamic properties of the system.
    \item Determine the thermodynamic functions for a photon gas confined to a $n$-dimensional torus (i.e. $X=\mathbb{T}^n$).
\end{itemize}
To our knowledge no comprehensive treatment of the various aspects of the Gribov problem in the finite temperature context has been given in
detail so far.\par

This paper is organized as follows: In Section 2 the geometrical structure of the space of thermal gauge fields is presented. It is shown that
there exist gauge transformations not connected to unity. As a consequence the bundle of thermal gauge fields over the space of gauge
inequivalent fields is not trivializable, which in physical terms is expressed by the statement that the theory suffers from Gribov
ambiguities.\par

Section 3 is devoted to the construction of a functional integral representation of the partition function in the space of thermal gauge
fields. In order to circumvent the Gribov problem we will apply a method which has been introduced some time ago in the stochastic quantization
scheme of Yang-Mills theory \cite{hüffel+keln1,hüffel+keln2}. For abelian gauge theories this procedure was elaborated recently in Ref.
\cite{kelnhofer}, where it was also explicitly shown why the conventional Faddeev-Popov procedure fails on non-simply connected manifolds.\par

The idea is to select a family of functional integral measures on the space of thermal gauge fields, whose domains of definition are determined
by local gauge fixing submanifolds and which are integrable along the orbits of the gauge group. Finally these local measures are glued
together in such a way that the physical relevant objects become independent of the chosen regularization of the gauge group and of the
particular way this gluing was provided. The redundant gauge degrees of freedom are taken into account by factoring out the regularized volume
of the full gauge group. This will be a slight extension of the original procedure \cite{hüffel+keln1}, where only those gauge transformations
were ruled out, which act freely on the space of thermal gauge fields. As a consequence of this reduction process, an additional topological
factor apart from the conventional Faddeev-Popov determinant will appear in the functional integral measure. Whereas this factor may be
neglected in the zero-temperature case, we will see that it contributes to the thermodynamical properties of the system and must be retained in
the finite temperature context. Additionally, different topological sectors may exist on a general manifold $X$. In order to include all
relevant thermal contributions, these sectors have to be considered as well when constructing the thermal partition function.\par

In accordance to the non-trivial topology of the space of inequivalent thermal gauge fields, the quantization is not unique. In the functional
integral approach this fact is usually taken into account by adding a total derivative to the classical Maxwell Lagrangian. This term does not
alter the classical equations of motion but contributes in the quantized version whenever the integration runs over topologically non-trivial
field configurations. In our case this additional action will be shown to be parametrized by a constant vector $\vec{\theta}\in
\mathbb{R}^{b_1(X)}$, where $b_1(X)$ is the first Betti number of $X$. The functional integral can be solved exactly and by using zeta function
regularization technique \cite{hawking, BVW, elizalde} we obtain a closed expression for the free energy of the system. This is achieved by
expressing the zeta-function of the Laplace operators on $\mathbb{T}_{\beta}^1\times X$ in terms of the zeta function associated with the
corresponding Laplace operators on $X$.\par

The quantum Maxwell theory at finite temperature is discussed from the canonical (Hamiltonian) point of view in Section 4. We will determine
the true phase space $\mathcal{P}$ and provide an explicit parametrization of this manifold. Since $\mathcal{P}$ turns out to be non-simply
connected, there exist inequivalent quantum theories which are classified by unitary irreducible representations of $\pi _1(\mathcal{P})$.
These are labelled by the non-integer values $\vec{\theta}\in \mathbb{R}^{b_1(X)}$. We will determine the Hamilton operator and calculate its
spectrum. The finite total vacuum energy of the system turns out to be the sum of the vacuum energy of the transverse gauge fields, which is
regularized by adopting the minimal subtraction scheme \cite{BVW}, and the energy of the $\theta$-states. The latter vanishes whenever
$\vec{\theta}\in\mathbb{Z}^{b_1(X)}$. The free energy is derived within the Hamiltonian scheme and the agreement with the functional integral
formalism is explicitly proved.

Finally, the high-temperature asymptotic expansion is calculated in terms of the heat kernel (Seeley) coefficients giving rise to finite size
and topological contributions to the familiar Stefan-Boltzmann black-body radiation law. \par

The scaling property of the system under constant scale transformations is analyzed in Section 5. Due to the regularization ambiguities of the
infinite vacuum energy and the non-trivial cohomology of $X$, the free energy does no longer transform homogenously. This leads to a modified
equation of state.\par

In Section 6 we discuss the example of finite temperature quantum Maxwell theory on the $n$-torus $X=\mathbb{T}^n$. The low- and high
temperature expansions are calculated for the main thermodynamic functions in terms of the Epstein zeta function \cite{Epstein1, Epstein2} and
the Riemann Theta function. Due to the $\theta$-term the ground state is degenerate. In the zero-temperature limit the entropy converges to the
logarithm of the degree of degeneracy, which proves explicitly the validity of the 3rd law of thermodynamics (i.e. Nernst theorem).\par

Necessary results on Riemann Theta functions and Epstein zeta functions are summarized in the Appendix. The conventions $c=\hbar =k_B=1$ are
used.\bigskip

\section{The configuration space of thermal gauge fields}

In this section we will consider the geometrical structure of the space of thermal gauge fields. These results are based on Ref.
\cite{kelnhofer}, where a detailed analysis can be found.\par

Geometrically, the finite temperature quantum Maxwell theory at equilibrium at finite temperature $1/\beta$ is regarded as gauge theory on the
product manifold $\mathbb{T}_{\beta}^1\times X$ with product metric $g =\gamma _{\beta}\oplus\gamma$. Here $\mathbb{T}_{\beta}^1$ denotes the
$1$-torus $\mathbb{T}^1$ equipped with the temperature dependent Riemannian metric $\gamma _{\beta}= \beta ^2dt\otimes dt$ ($t$ is the local
coordinate on $\mathbb{T}^1$) and the spatial background $X$ is a $n$-dimensional compact, connected, oriented and closed Riemannian manifold
equipped with a fixed metric $\gamma$. The corresponding volume forms on $\mathbb{T}_{\beta}^1$ and $X$ are denoted by
$vol_{\mathbb{T}_{\beta}^1}$ and $vol_{X}$, respectively.\par

Let $(Q,\pi _Q,\mathbb{T}_{\beta}^1\times X)$ be a principal $U(1)$-bundle over $\mathbb{T}_{\beta}^1\times X$ with projection $\pi _Q$. The
space of thermal gauge fields is identified with the $C^{\infty}$-Hilbert manifold $\mathcal{A}^{Q}$ of all connections of $Q$. The space of
$\mathbb{R}$-valued $k$-forms on $\mathbb{T}_{\beta}^1\times X$, denoted by $\Omega ^k(\mathbb{T}_{\beta}^1\times X)$, is equipped with the
$L^2$-inner product

\begin{equation}
<\upsilon _1,\upsilon _2>_{g}=\int _{\mathbb{T}_{\beta}^1\times X}\ \upsilon _1\wedge\star_{g}\ \upsilon _2\qquad\upsilon _1, \upsilon
_2\in\Omega ^k(\mathbb{T}_{\beta}^1\times X),
\end{equation}
where $\star_{g}$ is the Hodge star operator on $\mathbb{T}_{\beta}^1\times X$, satisfying $\star _{g}^2=(-1)^{k(n+1-k)}$ when acting on
$k$-forms. The co-differential $d_k^{\ast}=(-1)^{(n+1)(k+1)+1}\star _{g} d_{n+1-k}\star _{g}\colon\Omega ^k(\mathbb{T}_{\beta}^1\times
X)\rightarrow\Omega ^{k-1}(\mathbb{T}_{\beta}^1\times X)$ gives rise to the Laplacian operator $\Delta _k^{\mathbb{T}_{\beta}^1\times X}
=d_{k+1}^{\ast}d_k+d_{k-1}d_{k}^{\ast}$. Let $\mathcal{H}^k(\mathbb{T}_{\beta}^1\times X)$ denote the space of $\mathbb{R}$-valued harmonic
$k$-forms on $\mathbb{T}_{\beta}^1\times X$ and $\mathcal{H}^{k}(\mathbb{T}_{\beta}^1\times X)^{\perp}$ its orthogonal complement. Then we can
define the Green\rq s operator by

\begin{equation}
G_k^{\mathbb{T}_{\beta}^1\times X}\colon\Omega ^k(\mathbb{T}_{\beta}^1\times X)\rightarrow \mathcal{H}^{k}(\mathbb{T}_{\beta}^1\times
X)^{\perp},\quad G_k^{\mathbb{T}_{\beta}^1\times X}:= (\Delta _k^{\mathbb{T}_{\beta}^1\times X}\vert
_{\mathcal{H}^{k}(\mathbb{T}_{\beta}^1\times X)^{\perp}})^{-1}\circ \Pi ^{\mathcal{H}^{k}(\mathbb{T}_{\beta}^1\times X)^{\perp}},
\end{equation}
where $\Pi ^{\mathcal{H}^{k}(\mathbb{T}_{\beta}^1\times X)^{\perp}}$ is the projection onto $\mathcal{H}^{k}(\mathbb{T}_{\beta}^1\times
X)^{\perp}$. By construction one immediately obtains $\Delta _k^{\mathbb{T}_{\beta}^1\times X}\circ G_k^{\mathbb{T}_{\beta}^1\times
X}=G_k^{\mathbb{T}_{\beta}^1\times X}\circ\Delta _k^{\mathbb{T}_{\beta}^1\times X} =\Pi ^{\mathcal{H}^{k}(\mathbb{T}_{\beta}^1\times
X)^{\perp}}$. The dimension of $\mathcal{H}^k(\mathbb{T}_{\beta}^1\times X)$ is given by the $k$-th Betti number
$b_k(\mathbb{T}_{\beta}^1\times X)=b_k(X)+b_{k-1}(X)$. Here $b_k(X)$ is the $k$-th Betti number of $X$.\par

Let $\mathfrak{t}^1:=\sqrt{-1}\ \mathbb{R}$ denote the Lie-algebra of $U(1)$. The space of thermal gauge fields $\mathcal{A}^{Q}$ is an affine
space modelled on $\Omega ^1(\mathbb{T}_{\beta}^1\times X)\otimes\mathfrak{t}^1$ and can be equipped with the following metric

\begin{equation}
\hat{g}(w_1,w_2)=<\frac{w_1}{\sqrt{-1}},\frac{w_2}{\sqrt{-1}}>_{g}
\end{equation}
where $w_1, w_2\in\Omega ^1(\mathbb{T}_{\beta}^1\times X)\otimes\mathfrak{t}^1$. Let $vol_{\mathcal{A}^{Q}}^{\hat{g}}$ denote the induced
(formal) volume form on $\mathcal{A}^{Q}$.\par

The gauge group $\mathcal{G}$ is defined as the group of vertical bundle automorphisms on $Q$ and can be identified with the Hilbert Lie-Group
$C^{\infty}(\mathbb{T}_{\beta}^1\times X;U(1))$ of differentiable maps. The corresponding Lie algebra is $\mathfrak G=\Omega
^0(\mathbb{T}_{\beta}^1\times X;\mathfrak{t}^1)$. Since there is at least one non-contractible loop in $\mathbb{T}_{\beta}^1\times X$, the
gauge group contains gauge transformations which cannot be connected to the unity. These transformations are classified by a non-trivial $\pi
_0(\mathcal{G})$ and cannot be generated by taking the exponential of imaginary-valued functions on $\mathbb{T}_{\beta}^1\times X$. Under an
arbitrary gauge transformation $v\in\mathcal{G}$, the thermal gauge fields transform according to

\begin{equation}
A\mapsto A^v=A+(\pi _Q^{\ast}v)^{\ast}\vartheta ^{U(1)}\qquad v\in\mathcal{G},\label{action}
\end{equation}
where $\vartheta ^{U(1)}\in\Omega ^1(\mathbb{T}^1;\mathfrak{t}^1)$ is the Maurer Cartan form on $U(1)$. (For notational convenience we shall
not distinguish between $\pi _Q^{\ast}v$ and $v$.) The isotropy group of the non-free action \eqref{action} is $U(1)$, which corresponds to the
subgroup of constant gauge transformations. However, the quotient group $\mathcal{G}_{\ast}:=\mathcal{G}/U(1)$ provides a free action onto
$\mathcal{A}^{Q}$ giving rise to a smooth manifold $\mathcal{M}_{\ast}^{Q}=\mathcal{A}^{Q}/\mathcal{G}_{\ast}$ which represents the reduced
configuration space in the functional integral picture.\par

In order to fix notation, let $\mathcal{H}_{\mathbb{Z}}^k(X)$ denote the abelian group of harmonic $\mathbb{R}$-valued differential $k$-forms with
integer periods along $Z_k(X;\mathbb{Z})$, where $Z_k(X;\mathbb{Z})$ is the subcomplex of all closed smooth singular $k$-cycles on $X$ with
integer coefficients.\par

In our case we can choose a set of 1-cycles $c_i\in Z_1(X,\mathbb{Z})$, $i=1,\ldots ,b_1(X)$, whose corresponding homology classes $[c_i]$
provide a Betti basis of $H_1(X;\mathbb{Z})$. Let $\rho _j^{(n-1)}\in\mathcal{H}_{\mathbb{Z}}^{n-1}(X)$ be a basis associated to the Betti
basis $[c_i]$ via the Poincare duality, where $j=1,\ldots ,b_{n-1}(X)$. A dual basis $\rho _i^{(1)}\in\mathcal{H}_{\mathbb{Z}}^{1}(X)$ can be
adjusted such that $\int _{c_j}\rho _i^{(1)}=\int _{X}\rho _i^{(1)}\wedge\rho _j^{(n-1)}=\delta _{ij}$. Let $(h_{X}^{(1)})_{ij}=<\rho
_i^{(1)},\rho _j^{(1)}>_{\gamma}=\int _{X}\rho _i^{(1)}\wedge\star _{\gamma}\ \rho _j^{(1)}$ denote the induced metric on $\mathcal{H}^1(X)$.
Notice that $\star _{\gamma}\rho _i^{(1)}=\sum _{j=1}^{b_1(X)}(h_{X}^{(1)})_{ij}\rho _j^{(n-1)}$. The harmonic $1$-forms
$(pr_{\mathbb{T}_{\beta}^1}^{\ast}\varrho ^{(1)},pr_{X}^{\ast}\rho _1^{(1)},\ldots ,pr_{X}^{\ast}\rho _{b_1(X)}^{(1)})$ generate
$\mathcal{H}_{\mathbb{Z}}^1(\mathbb{T}_{\beta}^1\times X)$, where $\varrho ^{(1)}=\frac{1}{\beta}\ vol_{\mathbb{T}_{\beta}^1}$. Here
$pr_{\mathbb{T}_{\beta}^1}\colon \mathbb{T}_{\beta}^1\times X\rightarrow \mathbb{T}_{\beta}^1$ and $pr_{X}\colon \mathbb{T}_{\beta}^1\times
X\rightarrow X$ are the natural projections.\par

Let $(s_0,x_0)\in \mathbb{T}_{\beta}^1\times X$ be an arbitrary but fixed point. The gauge group $\mathcal{G}_{\ast}$ is equivalently
characterized as particular subgroup $\{v\in\mathcal{G}|v(s_0,x_0)=1\}$ of $\mathcal{G}$. The corresponding Lie algebra
$\mathfrak{G}_{\ast}=\mathfrak{G}/\mathfrak{t}^1$ contains all those $\xi\in\mathfrak G$ such that $\xi (s_0,x_0)=0$. We use this alternative
description to rewrite any (restricted) gauge transformation as product of an infinitesimal- and large gauge transformation. For this we
construct an isomorphism $\kappa :\mathfrak G_{\ast}\times\mathcal{H}_{\mathbb{Z}}^1(\mathbb{T}_{\beta}^1\times X)\rightarrow\mathcal{G}_{\ast}$ by

\begin{equation}
\kappa (\xi ,\alpha )(s,x)=\exp\xi (s,x)\cdot\exp{(2\pi\sqrt{-1}\int _{c_{(s,x)}}\alpha )}\label{gauge-group-split}.
\end{equation}
Here $c_{(s,x)}\colon [0,1]\rightarrow \mathbb{T}_{\beta}^1\times X$ is an arbitrary path in $\mathbb{T}_{\beta}^1\times X$ connecting
$(s_0,x_0)$ with $(s,x)$. As a result, \eqref{action} can be rewritten into the form

\begin{equation}
A\mapsto A+d_0\xi +2\pi\sqrt{-1}\left(m_0 pr_{\mathbb{T}_{\beta}^1}^{\ast}\varrho ^{(1)}+\sum _{j=1}^{b_1(X)}m_jpr_{X}^{\ast}\rho
_j^{(1)}\right),\quad\xi\in \mathfrak G_{\ast},\ m_0, m_j\in\mathbb{Z}.
\end{equation}
For any choice of an arbitrary but fixed background gauge field $A_0\in\mathcal{A}^{Q}$ there exists a smooth surjective map $\pi
_{\mathcal{M}_{\ast}^{Q}}^{A_0}\colon\mathcal{M}_{\ast}^{Q}\rightarrow\mathbb{T}^{1+b_1(X)}$, defined by

\begin{equation}
\pi _{\mathcal{M}_{\ast}^{Q}}^{A_0}([A])=\left( e^{\int _{\mathbb{T}_{\beta}^1\times \{x_0\}} (A-A_0)},e^{\int _{\{s_0\}\times c_{1}}
(A-A_0)}\ldots , e^{\int _{\{s_0\}\times c _{ b_1(X)}} (A-A_0)}\right).
\end{equation}
According to the general result proved in \cite{kelnhofer}, we thus obtain the following two propositions, which summarize the topological
structure of the space of thermal gauge fields:

\begin{prop}
$(\mathcal{A}^{Q},\pi _{\mathcal{A}^{Q}},\mathcal{M}_{\ast}^{Q})$ is a non trivializable flat principal $\mathcal{G}_{\ast}$-bundle over
$\mathcal{M}_{\ast}^{Q}$ with projection $\pi _{\mathcal{A}^{Q}}$. \qed
\end{prop}

\begin{prop}
For an arbitrary but fixed connection $A_0\in\mathcal{A}^{Q}$, $\pi
_{\mathcal{M}_{\ast}^{Q}}^{A_0}\colon\mathcal{M}_{\ast}^{Q}\rightarrow\mathbb{T}^{1+b_1(X)}$ admits the structure of a trivializable vector
bundle over $\mathbb{T}^{1+b_1(X)}$ with projection $\pi _{\mathcal{M}_{\ast}^{Q}}^{A_0}$ and typical fiber
$\mathcal{N}:=imd_2^{\ast}\otimes\mathfrak{t}^1$. If $A_0^{\prime}\in\mathcal{A}^{Q}$ is a different background connection, then $\pi
_{\mathcal{M}_{\ast}^{Q}}^{A_0^{\prime}}\colon\mathcal{M}_{\ast}^{Q}\rightarrow\mathbb{T}^{1+b_1(X)}$ is an isomorphic vector bundle. \qed
\end{prop}
As a result, the manifolds $\mathcal{A}^{Q}$ and $\mathbb{T}^{1+b_1(X)}\times\mathcal{N}\times\mathcal{G}_{\ast}$ are locally diffeomorphic. An
explicit expression for these local diffeomorphisms can be given as follows: Let us introduce the two contractible open sets $V_{a_j=1}=
\mathbb{T}^1\backslash\{northern pole\}$ and $V_{a_j=2}= \mathbb{T}^1\backslash\{southern pole\}$, which cover the $j$-th 1-torus
$\mathbb{T}^1\subset\mathbb{T}^{1+b_1(X)}$. Then

\begin{equation}
\mathcal{V}=\{V_a:=V_{a_0}\times\cdots\times V_{a_j}\times\cdots\times V_{a_{b_1(X)}}\vert\quad a:=(a_0,\ldots ,a_j,\ldots ,a_{b_1(X)}),\quad
a_{j}\in\{1,2\}\},
\end{equation}
is an open cover of $\mathbb{T}^{1+b_1(X)}$. Hence the family of open sets $U_a^{A_0}:=(\pi_{\mathcal{M}_{\ast}^{Q}}^{A_0})^{-1}(V_a)$ provides
a finite open cover $\mathcal{U}^{A_0}=\{U_a^{A_0}\}$ of $\mathcal{M}_{\ast}^{Q}$. Any two local sections $\sigma_{a_{j}}:V_{a_j}\rightarrow
\mathbb{R}^1$ of the universal covering bundle $\mathbb{R}^1 \rightarrow \mathbb{T}^1$ generate $2^{1+b_1(X)}$ local sections $\sigma_a\colon
V_a\subset \mathbb{T}^{1+b_1(X)}\rightarrow\mathbb{R}^{1+b_1(X)}$, given by $\sigma_a=(\sigma_{a_0},\cdots ,\sigma_{a_{b_1(X)}})$. It can be
shown that the maps $\chi _a^{A_0}\colon V_a\times\mathcal{N}\times\mathcal{G}_{\ast}\rightarrow
\pi_{\mathcal{A}^{Q}}^{-1}(U_a^{A_0})\subseteq\mathcal{A}^{Q}$, defined by

\begin{equation}
\chi _a^{A_0}(z_0,\ldots , z_{b_1(X)},\tau ,v)=A_0+2\pi\sqrt{-1}\left( \sigma_{a_0}(z_0)\ pr_{\mathbb{T}_{\beta}^1}^{\ast}\varrho +\sum
_{i=1}^{b_1(X)}\sigma_{a_i}(z_i)\ pr_{X}^{\ast}\rho _i^{(1)}\right) +\tau +v^{\ast}\vartheta ^{U(1)}.\label{diffeo}
\end{equation}
provide an appropriate family of local diffeomorphisms.

\section{The free energy in the functional integral approach}

\subsection{The construction principle}

In this section we want to construct a functional integral representation of the (thermal) partition function for the photon gas. The Maxwell
fields are governed by the classical action

\begin{equation}
S_{inv}(A)=\frac{1}{2}\|F_A\|^2=\frac{1}{2}\ \hat g(F_A,F_A).
\end{equation}
In the previous section it has been shown that the reduced configuration space is the non-simply connected manifold $\mathcal{M}_{\ast}^{Q}$,
since $\pi _{1}(\mathcal{M}_{\ast}^{Q})=\pi _{0}(\mathcal{G}_{\ast})\cong \mathbb{Z}^{1+b_1(X)}$. It is well known that ambiguities arise in
the definition of the vacuum if the corresponding configuration space is non-simply connected. In the functional integral approach this fact is
usually taken into account by adding a so-called topological- or theta action, denoted by $S_{\theta}$, to the gauge invariant classical
action.
\par

To each arbitrary but fixed vector $\vec{\theta}=(\theta _1,\ldots ,\theta _{b_1(X)})\in\mathbb{R}^{b_1(X)}$ we associate the harmonic $1$-form
$\underline{\theta}:=2\pi\sqrt{-1}\sum _{i,j=1}^{b_1(X)}(h_{X}^{(1)})_{ij}^{-1}\theta _i\rho
_j^{(1)}\in\mathcal{H}^{1}(X)\otimes\mathfrak{t}^1$. We propose the following action

\begin{equation}
\begin{split}
S_{\theta}(A):&=\frac{1}{2\pi\sqrt{-1}}\ \hat g(F_A,\star _gpr_{X}^{\ast}\star _{\gamma}\underline{\theta})\\
&=-\sum_{i=1}^{b_1(X)}\theta _i\int_{\mathbb{T}_{\beta}^1\times X}F_A\wedge pr_{X}^{\ast}\rho _i^{(n-1)},\label{theta-action}
\end{split}
\end{equation}
which is indeed topological since it is independent of the metric of $\mathbb{T}_{\beta}^1\times X$. This specific choice will become more
transparent in the next section when the Hamiltonian approach is considered. For the total action we take now $S_{tot}:=S_{inv}+S_{\theta}$
instead of $S_{inv}$ alone. Since $\frac{\delta S_{\theta}}{\delta A}=0$, the inclusion of this additional term does not change the classical
equations of motion. The theta action $S_{\theta}$ is imaginary because our context is the finite-temperature Euclidean regime where time is
compactified. In fact, analytic continuation of \eqref{theta-action} onto the real time axis is necessary to obtain the correct real-valued
theta-term in the Hamiltonian formulation on space-time $\mathbb{R}\times X$.\par

The total configuration space, denoted by $\mathcal{A}_{\mathbb{T}_{\beta}^1\times X}$, is the disjoint union

\begin{equation}
\mathcal{A}_{\mathbb{T}_{\beta}^1\times X}=\bigsqcup _{Q}\mathcal{A}^{Q}
\end{equation}
over equivalence classes of principal $U(1)$-bundles $Q$ over the base manifold $\mathbb{T}_{\beta}^1\times X$. Each class is labelled by its
first Chern-class $c_1(Q)\in H^2(\mathbb{T}_{\beta}^1\times X;\mathbb{Z})\cong H^1(X;\mathbb{Z})\oplus H^2(X;\mathbb{Z})$.\par

In the conventional functional integral formulation the integration of the functional $e^{-S_{tot}}vol_{\mathcal{A}^{Q}}^{\hat{g}}$ over
$\mathcal{A}^{Q}$ would become infinite due to the gauge invariance of the classical action. Our aim is to give this integral a well-defined
meaning by damping the contributions from the gauge dependent degrees of freedom. The non-trivial bundle structure
$\mathcal{A}^{Q}\xrightarrow{\pi _{\mathcal{A}^{Q}}}\mathcal{M}_{\ast}^{Q}$ allows only for a local damping procedure, which, however, can be
patched together using a partition of unity in the end.\par

The map $[\xi]\mapsto d_1^{\ast}d_0G_0^{\mathbb{T}_{\beta}^1\times X}(\xi )$ provides an isomorphism $\mathfrak G_{\ast}\cong
imd_1^{\ast}\otimes\mathfrak t^1$. Notice that $\mathcal{G}$ is a trivializable principal $U(1)$-bundle over $\mathcal{G}_{\ast}$, whose global
trivialization is given by the diffeomorphism $\hat{\sigma} (u)=(uu(s_0,x_0)^{-1}, u(s_0,x_0))$ where $u\in\mathcal{G}$. Let $\vartheta
^{\mathcal{G}}\in\Omega ^1(\mathcal{G})\otimes\mathfrak G$ and $\vartheta ^{\mathcal{G}_{\ast}}\in\Omega ^1(\mathcal{G}_{\ast})\otimes\mathfrak
G_{\ast}$ denote the Maurer Cartan forms on $\mathcal{G}$ and $\mathcal{G}_{\ast}$, respectively. Formally one can introduce the natural
metrics $\Upsilon ^{\mathcal{G}}=\hat{g}(\vartheta ^{\mathcal{G}}(.),\vartheta ^{\mathcal{G}}(.))$ and $\Upsilon
^{\mathcal{G}_{\ast}}=\hat{g}(\vartheta ^{\mathcal{G}_{\ast}}(.),\vartheta ^{\mathcal{G}_{\ast}}(.))$ on these two gauge groups. Let
$vol_{\mathcal{G}}$ and $vol_{\mathcal{G}_{\ast}}$ denote the induced left-invariant volume forms. Furthermore $U(1)$ is equipped with the
standard flat metric giving rise to the volume form $vol_{U(1)}=(\sqrt{-1})^{-1}\vartheta ^{U(1)}$. With respect to our parametrization of
$U(1)$, the corresponding volume is $\int _{U(1)}vol_{U(1)}=2\pi$.\par

Let us now introduce three real-valued regularizing functions $\Lambda _{reg}^{\mathcal{G}}\in C^{\infty}(\mathcal{G})$, $\Lambda
_{reg}^{\mathcal{G}_{\ast}}\in C^{\infty}(\mathcal{G}_{\ast})$ and $\Lambda _{reg}^{U(1)}\in C^{\infty}(U(1))$ in such a way that the following
correspondingly regularized group volumes

\begin{equation}
\begin{split}
& Vol(\mathcal{G};e^{-\Lambda _{reg}^{\mathcal{G}}}):=\int _{\mathcal{G}}\ vol_{\mathcal{G}}\ e^{-\Lambda _{reg}^{\mathcal{G}}} <\infty \\
& Vol(\mathcal{G}_{\ast};e^{-\Lambda _{reg}^{\mathcal{G}_{\ast}}}):=\int _{\mathcal{G}_{\ast}}\ vol_{\mathcal{G}_{\ast}}\ e^{-\Lambda
_{reg}^{\mathcal{G}_{\ast}}} <\infty\\ &Vol(U(1);e^{-\Lambda _{reg}^{U(1)}}):=\int _{U(1)}\ vol_{U(1)}\ e^{-\Lambda
_{reg}^{U(1)}}\end{split}\label{group-volume-1}
\end{equation}
become finite. Evidently, a regularization of $U(1)$ is not necessary (of course $\Lambda _{reg}^{U(1)}=0$ would be sufficient) but we will use
this freedom later on to absorb irrelevant multiplicative factors in the partition function. \par

In order to relate these volumes, we consider the differential of $\hat{\sigma} ^{-1}$, which gives

\begin{equation}
T\hat{\sigma} ^{-1}(\tilde{\xi},w)=Tr_{z}^{\mathcal{G}}\left( \tilde{\xi}+\frac{d}{dt}|_{t=0}\ v\cdot e^{t\vartheta
^{U(1)}(w)}\right),\qquad\tilde{\xi}\in T_v\mathcal{G}_{\ast},\ w\in T_zU(1),
\end{equation}
where $r_{z}^{\mathcal{G}}$ denotes the right multiplication on $\mathcal{G}$ by $z\in U(1)$. Since $\vartheta
^{\mathcal{G}}(\tilde{\xi})=\vartheta ^{\mathcal{G}_{\ast}}(\tilde{\xi})\in imd_{1}^{\ast}\otimes\mathfrak t^1$ the induced metrics are related
by

\begin{equation}
(\hat{\sigma} ^{-1})^{\ast}\Upsilon ^{\mathcal{G}}=\Upsilon ^{\mathcal{G}_{\ast}}-\beta V\ \vartheta ^{U(1)}(.)\vartheta ^{U(1)}(.),
\end{equation}
where $V:=\int _{X}vol_{X}$ is the volume of $X$ with respect to the metric $\gamma$. For the volume forms, this finally implies

\begin{equation}
(\hat{\sigma} ^{-1})^{\ast}vol_{\mathcal{G}}=(\beta V)^{\frac{1}{2}}\ vol_{\mathcal{G}_{\ast}}\wedge vol_{U(1)}.\label{group-volume-relation}
\end{equation}
Let us take the natural choice $\Lambda _{reg}^{\mathcal{G}}:=\pi _{\mathcal{G}}^{\ast}\Lambda
_{reg}^{\mathcal{G}_{\ast}}+\hat{\sigma}_{(s_0,x_0)}^{\ast}\Lambda _{reg}^{U(1)}$, where $\pi
_{\mathcal{G}}:\mathcal{G}\rightarrow\mathcal{G}_{\ast}$ is the projection and $\hat{\sigma}_{(s_0,x_0)}(u):=u(s_0,x_0)$. Using
\eqref{group-volume-1} it follows that

\begin{equation}
Vol(\mathcal{G};e^{-\Lambda _{reg}^{\mathcal{G}}})= (\beta V)^{\frac{1}{2}}\ Vol(\mathcal{G}_{\ast};e^{-\Lambda
_{reg}^{\mathcal{G}_{\ast}}})Vol(U(1);e^{-\Lambda _{reg}^{U(1)}}).\label{group-volume-2}
\end{equation}
An appropriate choice for the regularizing function is provided by

\begin{equation}
\Lambda _{reg}^{\mathcal{G}_{\ast}}(v)=\frac{1}{2}\|d_1^{\ast}(v^{\ast}\vartheta ^{U(1)})\|^2+\frac{1}{2}\|\Pi
^{\mathcal{H}^{1}(\mathbb{T}_{\beta}^1\times X)}(v^{\ast}\vartheta ^{U(1)})\|^2,\label{volume-gauge-concrete}
\end{equation}
where $\Pi ^{\mathcal{H}^{1}(\mathbb{T}_{\beta}^1\times X)}$ is the projector onto $\mathcal{H}^{1}(\mathbb{T}_{\beta}^1\times X)$
\cite{kelnhofer}. According to \eqref{gauge-group-split} and with respect to the chosen Betti-basis, each $v\in\mathcal{G}_{\ast}$ is uniquely
represented by a pair $(\xi ,\vec{m})\in (im d_1^{\ast}\otimes\mathfrak t^1)\times \mathbb{Z}^{1+b_1(X)}$. As a consequence the integration
over $\mathcal{G}_{\ast}$ splits into an integral over $im d_1^{\ast}\otimes\mathfrak t^1$ and a summation over the components of the vector
$\vec{m}=(m_{0},\ldots ,m_{b_1(X)})\in\mathbb{Z}^{1+b_1(X)}$. For the choice \eqref{volume-gauge-concrete} one obtains

\begin{equation}
Vol(\mathcal{G}_{\ast};e^{-\Lambda _{reg}^{\mathcal{G}_{\ast}}})= (\det{\Delta _0^{\mathbb{T}_{\beta}^1\times
X}|_{\mathcal{H}^0(\mathbb{T}_{\beta}^1\times X)^{\perp}}})^{-1}\Theta _{b_1(X)}(0|2\pi\sqrt{-1}h_{\mathbb{T}_{\beta}^1\times X}^{(1)}),
\end{equation}
where $\Theta _{b_1(X)}(.|.)$ denotes the $b_1(X)$-dimensional Riemann-Theta function \eqref{theta-original}.\par

Now we come to the construction of the functional integral representation for the thermal partition function. Due to the non-trivial bundle
structure of the space of thermal gauge fields the separation in gauge independent and gauge dependent degrees of freedom can be done only
locally. The idea is to construct integrable measures locally in $\mathcal{A}$ and in the end to paste them together with a partition of unity
to obtain a global and integrable measures. This method was originally introduced for studying the relation between stochastic quantization and
the conventional Faddeev-Popov quantization scheme \cite{hüffel+keln1} and then applied to define an integrable partition function for
Yang-Mills theory in order to overcome the Gribov problem \cite{hüffel+keln2}. Recently this approach was used for the formulation of an
appropriate functional integral representation of generalized $p$-form gauge fields \cite{keln3}.\par

Let $\{p_a\}$ be a partition of unity of $\mathcal{M}_{\ast}^{Q}$ subordinate to $\mathcal{U}^{A_0}$ and define $\omega
_a^{A_0}:=pr_{\mathcal{G}_{\ast}}\circ\varphi _a^{A_0}$, where $\varphi _a^{A_0}\colon\pi_{\mathcal{A}^{Q}}^{-1}(U_a)\rightarrow
U_a\times\mathcal{G}_{\ast}$ is a family of local trivializations of the bundle $\mathcal{A}^{Q}\xrightarrow{\pi
_{\mathcal{A}^{Q}}}\mathcal{M}_{\ast}^{Q}$ and $pr_{\mathcal{G}_{\ast}}\colon U_a^{A_0}\times\mathcal{G}_{\ast}\rightarrow\mathcal{G}_{\ast}$
is the natural projection. Let us introduce the following smooth functional on $\mathcal{A}^{Q}$,

\begin{equation}
\Xi ^{Q}=\sum _{a}(\pi _{\mathcal{A}^{Q}}^{\ast}p_a)\ e^{-(\omega _a^{A_0})^{\ast}\Lambda
_{reg}^{\mathcal{G}_{\ast}}},\label{additional-factor}
\end{equation}
which accounts for the regularized group volume $\mathcal{G}_{\ast}$. Now we define the functional integral representation for the thermal
partition function by

\begin{equation}
\mathcal{Z}(\beta ,V)= N\ \sum _{c_1(Q)\in H^2(\mathbb{T}_{\beta}^1\times X;\mathbb{Z})}\int _{\mathcal{A}^{Q}}\
\frac{vol_{\mathcal{A}^{Q}}^{\hat{g}}}{Vol(\mathcal{G};e^{-\Lambda _{reg}^{\mathcal{G}}})}\ \Xi ^{Q}\ e^{-S_{tot}}.\label{partition-total}
\end{equation}
The partition function is the formal sum over all equivalence classes of principal $U(1)$ bundles over $\mathbb{T}_{\beta}^1\times X$
classified by their first Chern-classes $c_1(Q)$ and the functional integration over the corresponding spaces $\mathcal{A}^{Q}$ of thermal
gauge fields. The normalization constant $N$ is temperature- and volume independent and will be fixed later on. The volume of the total gauge
group $\mathcal{G}$ is factored out in order to reduce the system to the true physical degrees of freedom. Notice that this is a slight
generalization of the original procedure introduced in Ref. \cite{hüffel+keln2}, where only the volume of the restricted gauge group
$\mathcal{G}_{\ast}$ was factored out. It will be shown that this proposed partition function is integrable and resolves the Gribov problem.
Moreover, the phase $e^{-S_{\theta}}$ gives the weight-factor for the different topological sectors.\par

Physical observables are regarded as gauge invariant functions on $\mathcal{A}_{\mathbb{T}_{\beta}^1\times X}$. The thermal expectation value
(TEV) of a physical observable $f\in C^{\infty}(\mathcal{A}^Q)$ is defined by

\begin{equation}
<f>_{\beta}=\frac{\sum _{c(Q)\in H^2(\mathbb{T}_{\beta}^1\times X;\mathbb{Z})}I_{\beta}^{Q}(f)}{\mathcal{Z}(\beta ,V)},\label{VEV-1}
\end{equation}
where

\begin{equation}
I_{\beta}^{Q}(f)=\int _{\mathcal{A}^{Q}}\ \frac{vol_{\mathcal{A}^{Q}}^{\hat{g}}}{Vol(\mathcal{G};e^{-\Lambda _{reg}^{\mathcal{G}}})}\
e^{-S_{tot}(\beta )}\ \Xi ^{Q}\ f.\label{VEV-2}
\end{equation}
As a consequence the TEV of a physical observable is independent of the particular choices of the regularization, the local trivializations and
the partition of unity (for an explicit proof see \cite{hüffel+keln2}).\par

\subsection{Determination of the free energy}

In accordance with the bundle structure of $\mathcal{A}^Q$, the functional integral in \eqref{partition-total} can be transformed into an
integral over $\mathbb{T}^{1+b_1(X)}\times\mathcal{N}\times\mathcal{G}_{\ast}$, which can be explicitly calculated. Using the family of local
diffeomorphisms \eqref{diffeo}, one obtains for the transformed measure
\begin{equation}
(\chi _a^{A_0})^{\ast}vol_{\mathcal{A}^{Q}}^{\hat{g}}= \left(\det h_{\mathbb{T}_{\beta}^1\times X}^{(1)}\ \det{\Delta
_0^{\mathbb{T}_{\beta}^1\times X}|_{\mathcal{H}^0(\mathbb{T}_{\beta}^1\times X)^{\perp}}}\right)^{\frac{1}{2}}\ vol_{
\mathbb{T}^{1+b_1(X)}}\vert _{V_a}\wedge vol_{\mathcal{N}}\wedge vol_{\mathcal{G}_{\ast}},\label{volume-form}
\end{equation}
where $vol_{ \mathbb{T}^{1+b_1(X)}}\vert _{V_a}$ is the induced volume form on $\mathbb{T}^{1+b_1(X)}$, yet restricted to the patch $V_a$. It
follows that $vol_{\mathbb{T}^{1+b_1(X)}}=\hat{i}_{0}^{\ast}vol_{U(1)}\wedge\ldots \wedge\hat{i}_{b_1(X)}^{\ast}vol_{U(1)}$ where
$\hat{i}_{k}\colon U(1)\hookrightarrow\mathbb{T}^{1+b_1(X)}$ denotes the $k$-th inclusion for $k=0,\ldots ,b_1(X)$. The restriction of
$\hat{g}$ to $\mathcal{N}$ induces a flat metric on that space and an associated volume form $vol_{\mathcal{N}}$. Apart from the conventional
Faddeev-Popov determinant of the scalar Laplacian, the factor $\det h_{\mathbb{T}_{\beta}^1\times X}^{(1)}$ appears in addition in the induced
measure \eqref{volume-form}. This factor carries the topological information of $\mathbb{T}_{\beta}^1\times X$ and will be shown to depend on
temperature and volume. On the contrary to the zero-temperature case, where this factor as well as the conventional Faddeev-Popov determinant
may be neglected, it is essential to keep both factors in the functional integral in the finite temperature context. \par

A direct calculation gives

\begin{equation}
\begin{split}
&\star _{g}\ \left( pr_{\mathbb{T}_{\beta}^1}^{\ast} \varrho ^{(1)}\right)=\frac{1}{\beta} \ pr_{X}^{\ast} vol_X,\\
&\star _{g}\ \left( pr_{X}^{\ast} \rho _j^{(1)} \right)=-pr_{\mathbb{T}_{\beta}^1}^{\ast}\ vol_{\mathbb{T}_{\beta}^1}\wedge pr_{X}^{\ast}(\star
_{\gamma}\rho _j^{(1)}),
\end{split}
\end{equation}
so that the induced metric on $\mathcal{H}^1(\mathbb{T}_{\beta}^1\times X)$ admits the following matrix of rank $1+b_1(X)$,

\begin{equation}
h_{\mathbb{T}_{\beta}^1\times X}^{(1)}=\begin{pmatrix}
  \beta^{-1}V & 0 \\
  0 & \beta h_{X}^{(1)}
\end{pmatrix}.\label{metric-0}
\end{equation}
Hence

\begin{equation}
\det{h_{\mathbb{T}_{\beta}^1\times X}^{(1)}}=\beta ^{b_1(X)-1}V\det{h_{X}^{(1)}}.\label{metric-1}
\end{equation}
Due to the Hodge decomposition theorem it is always possible to choose the fixed background gauge field $A_0\in\mathcal{A}^{Q}$ in such a way
that the corresponding field strength $F_{A_0}$ becomes harmonic, i.e.
$\frac{1}{2\pi\sqrt{-1}}F_{A_0}\in\mathcal{H}_{\mathbb{Z}}^2(\mathbb{T}_{\beta}^1\times X)$. Since the cohomology of $X$ is finitely generated,
the Chern-class $c_1(Q)$ admits the following (non-canonical) decomposition

\begin{equation}
c_1(Q)=\left( \sum _{i=1}^{b_1(X)}l_i \eta _{i}^{(1)},\sum _{j=1}^{b_2(X)}m_{j} \eta _{j}^{(2)}+ \sum _{k=1}^{w}\ y_{k}t_{k}^{(2)}\right)\in
H^1(X;\mathbb{Z})\oplus H^2(X;\mathbb{Z}),\label{chern-class}
\end{equation}
with respect to the Betti bases $(\eta _{i}^{(1)})_{i=1}^{b_1(X)}$ and $(\eta _{j}^{(2)})_{j=1}^{b_2(X)}$ for $H^1(X;\mathbb{Z})$ and
$H^2(X;\mathbb{Z})$, respectively. Here $b_2(X)=\dim H^2(X;\mathbb{R})$. The torsion subgroup $TorH^2(X;\mathbb{Z})$ is generated by the basis
$(t_{k}^{(2)})_{k=1}^{w}$. Hence there exists $r_1,\ldots , r_w\in\mathbb{N}$ such that $r_kt_{k}^{(2)}=0$ for each $k=1,\ldots ,w$. The order
of the torsion subgroup is $|TorH^2(X;\mathbb{Z})|=\prod_{k=1}^{w}r_k$. Finally the vectors $(l_1,\ldots ,l_{b_1(X)})\in\mathbb{Z}^{b_1(X)}$,
$(m_1,\ldots ,m_{b_2(X)})\in\mathbb{Z}^{b_2(X)}$ and $y_{k}\in \mathbb{Z}_{r_k}\equiv \mathbb{Z}/r_k\mathbb{Z}$ denote the components with
respect to these different generators.\par

Let $(\rho _j^{(2)})_{j=1}^{b_2(X)}$ be generators for $\mathcal{H}_{\mathbb{Z}}^2(X)$, then $(pr_{\mathbb{T}_{\beta}^1}^{\ast}\varrho
^{(1)}\wedge pr_{X}^{\ast}\rho _i^{(1)})_{i=1}^{b_1(X)}$ and $(pr_{X}^{\ast}\rho _j^{(2)})_{j=1}^{b_2(X)}$ are the induced generators for
$\mathcal{H}_{\mathbb{Z}}^2(\mathbb{T}_{\beta}^1\times X)$. The background field strength can then be expressed in the form

\begin{equation}
F_{A_0}=2\pi\sqrt{-1}\left[\sum _{i=1}^{b_1(X)}l_i\ pr_{\mathbb{T}_{\beta}^1}^{\ast}\varrho ^{(1)}\wedge pr_{X}^{\ast}\rho _i^{(1)}+\sum
_{j=1}^{b_2(X)}m_j\ \ pr_X^{\ast}\ \rho _j^{(2)}\right].\label{field-strength}
\end{equation}
A direct calculation gives

\begin{equation}
\begin{split}
&\star _{g} \left(pr_X^{\ast}\ \rho _j^{(2)}\right) =\beta\ pr_{\mathbb{T}_{\beta}^1}^{\ast}\varrho ^{(1)}\wedge pr_X^{\ast}\left( \star
_{\gamma}\
\rho _j^{(2)}\right)\\
&\star _{g}\left(pr_{\mathbb{T}_{\beta}^1}^{\ast}\varrho ^{(1)}\wedge pr_{X}^{\ast}\rho _i^{(1)})\right) =\frac{1}{\beta}\
pr_X^{\ast}\left(\star _{\gamma}\rho _i^{(1)}\right),
\end{split}
\end{equation}
yielding the following metric $(h_{\mathbb{T}_{\beta}^1\times X}^{(2)})$ on $\mathcal{H}^2(\mathbb{T}_{\beta}^1\times X)$, namely
\begin{equation}
h_{\mathbb{T}_{\beta}^1\times X}^{(2)}=\begin{pmatrix}
  \beta^{-1}h_{X}^{(1)} & 0 \\
  0 & \beta h_{X}^{(2)}
\end{pmatrix},\label{metric-2}
\end{equation}
of rank $b_1(X)+b_2(X)$. Here $(h_{X}^{(2)})_{ij}=<\rho _i^{(2)},\rho _j^{(2)}>_{\gamma}$ is the induced metric on $\mathcal{H}^2(X)$.\par

Let us recall that on an arbitrary compact manifold $M$ the Laplace operator, when restricted to $\mathcal{H}^k(M)^{\perp}$, splits into the
sum $\Delta _k^{M}|_{\mathcal{H}^k(M)^{\perp}}=\Delta _k^{M}|_{im d_{k-1}}+\Delta _k^{M}|_{im d_{k+1}^{\ast}}$. Since the spectra of $\Delta
_k^{M}|_{im d_{k-1}}$ and $\Delta _{k-1}^{M}|_{im d_{k}^{\ast}}$ coincide, the spectrum of $\Delta _k^{M}|_{\mathcal{H}^k(M)^{\perp}}$ is the
union of eigenvalues of $\Delta _k^{M}|_{im d_{k+1}^{\ast}}$ and of $\Delta _{k-1}^{M}|_{im d_{k}^{\ast}}$. In our case
$M=\mathbb{T}_{\beta}^1\times X$. This implies for the determinant

\begin{equation}
\det{\Delta _1^{\mathbb{T}_{\beta}^1\times X}|_{\mathcal{H}^1(\mathbb{T}_{\beta}^1\times X)^{\perp}}}=(\det{\Delta
_1^{\mathbb{T}_{\beta}^1\times X}|_{im d_2^{\ast}}})\ (\det{\Delta _0^{\mathbb{T}_{\beta}^1\times X}|_{\mathcal{H}^0(\mathbb{T}_{\beta}^1\times
X)^{\perp}}}).\label{determinants}
\end{equation}
Using \eqref{group-volume-relation}, \eqref{volume-form}, \eqref{metric-1}, \eqref{field-strength}, \eqref{metric-2} and \eqref{determinants},
the integration in \eqref{VEV-2} can be carried out and gives (for $f=1$)

\begin{equation}
\begin{split}
I_{\beta}^{Q}(1)= &(2\pi )^{(1+b_1(X))}\ \beta ^{\frac{b_1(X)-2}{2}}\ (\det h_X^{(1)})^{\frac{1}{2}}\ (\det{\Delta
_0^{\mathbb{T}_{\beta}^1\times X}|_{\mathcal{H}^0(\mathbb{T}_{\beta}^1\times X)^{\perp}}}) (\det{\Delta _1^{\mathbb{T}_{\beta}^1\times
X}|_{\mathcal{H}^1(\mathbb{T}_{\beta}^1\times X)^{\perp}}})^{-\frac{1}{2}}\\ &\times \exp{\left(-\frac{(2\pi)^2}{2}\left[\beta ^{-1}\sum
_{i,j=1}^{b_1(X)}(h_X^{(1)})_{ij}l_{i}l_{j}+ \beta\sum _{i,j=1}^{b_2(X)}(h_X^{(2)})_{ij}m_im_j\right]+2\pi\sqrt{-1}\sum _{i=1}^{b_1(X)}\theta
_il_i\right)}\\ &\times Vol(U(1);e^{-\Lambda _{reg}^{U(1)}})^{-1} N.\label{partition-monopole}
\end{split}
\end{equation}
The pair of vectors $(\vec{l},\vec{m})\in\mathbb{Z}^{b_1(X)}\times\mathbb{Z}^{b_2(X)}$ labels the inequivalent principal $U(1)$-bundles $Q$.
Using the decomposition \eqref{chern-class}, the sum over the Chern-classes $c_1(Q)$ is performed by summing \eqref{partition-monopole} over
the free part and the torsion part of $H^2(\mathbb{T}_{\beta}^1\times X;\mathbb{Z})$, respectively.\par

Let us choose $e^{-\Lambda _{reg}^{U(1)}}:=(2\pi)^{\frac{b_1(X)}{2}}$ for the regularizing functional on $U(1)$, then the thermal partition
function \eqref{partition-total} admits the following form

\begin{equation}
\begin{split}
\mathcal{Z}^{\theta}(\beta ,V)= &(2\pi)^{\frac{b_1(X)}{2}}\beta ^{\frac{b_1(X)-2}{2}}\ (\det h_X^{(1)})^{\frac{1}{2}}\ (\det{\Delta
_0^{\mathbb{T}_{\beta}^1\times X}|_{\mathcal{H}^0(\mathbb{T}_{\beta}^1\times X)^{\perp}}})\ (\det{\Delta _1^{\mathbb{T}_{\beta}^1\times X}
|_{\mathcal{H}^1(\mathbb{T}_{\beta}^1\times X)^{\perp}}})^{-\frac{1}{2}}\\
&\times\Theta _{b_1(X)}\left(\vec{\theta} |2\pi\sqrt{-1}\ \beta ^{-1}\ h_X^{(1)}\right) \Theta _{b_2(X)}\left(0 |2\pi\sqrt{-1}\ \beta\
h_X^{(2)}\right) |TorH^2(X;\mathbb{Z})| N.\label{partition-final}
\end{split}
\end{equation}
In order to give the formal determinants arising in \eqref{partition-final} a mathematical meaning, the zeta-function method will be used as
regularization technique. Generally, the regularized determinant of a non-negative, self-adjoint elliptic operator $\mathcal{D}$ of second
order on a general manifold $M$ is defined by

\begin{equation}
\det \mathcal{D} =\exp{\left( -\zeta ^{\prime} (0;\mathcal{D})\right)}\equiv\exp{\left( -\frac{d}{ds}|_{s=0}\zeta (s;\mathcal{D})\right)}.
\end{equation}
Here $\zeta (s;\mathcal{D})$ denotes the zeta-function of the operator $\mathcal{D}$, defined by

\begin{equation}
\zeta (s;\mathcal{D})=\sum _{\nu _{\alpha}(\mathcal{D})\neq 0}\nu _{\alpha}(\mathcal{D})^{-s}=\frac{1}{\Gamma (s)}\int _0^{\infty}dt\ t^{s-1}
Tr(e^{-t\mathcal{D}}-\Pi ^{\mathcal{D}}),\quad s\in\mathbb{C}\label{zeta}
\end{equation}
where the sum runs over the non vanishing eigenvalues $\nu _{\alpha}(\mathcal{D})$ of $\mathcal{D}$ only. In this sum each eigenvalue appears
the same number of times as its multiplicity. This sum is convergent for $\Re(s)>\frac{\dim M}{2}$. The second equation in \eqref{zeta} is the
heat-kernel representation of the zeta-function based on the Mellin transformation. Therein $\Pi ^{\mathcal{D}}$ is the projector onto the
kernel of $\mathcal{D}$. Let $\mathcal{D}'$ denote the restriction of $\mathcal{D}$ to the non-zero modes, i.e.
$\mathcal{D}':=\mathcal{D}|_{\ker \mathcal{D}^{\perp}}$, then $\zeta (s;\mathcal{D})=\zeta (s;\mathcal{D}')$ holds by construction. In general,
the operator $\mathcal{D}$ possesses the following asymptotic expansion

\begin{equation}
Tr(e^{-t\mathcal{D}})\simeq\ \sum _{k=0}^{\infty}a_k(\mathcal{D})\ t^{\frac{k-\dim M}{2}},\quad\textrm{for}\ t\downarrow 0,\label{seeley1}
\end{equation}
with respect to the asymptotic sequence $t\mapsto t^{\frac{k-\dim M}{2}}$ of functions \cite{Gilkey, Kirsten, Gilkey2}. The constants
$a_k(\mathcal{D})$ are the Seeley coefficients of $\mathcal{D}$. By using this expansion and splitting the integral in \eqref{zeta} into an
integral over $[0,1]$ and $[1,\infty )$ respectively, the zeta-function admits the asymptotic expansion

\begin{equation}
\zeta (s;\mathcal{D})\simeq\frac{1}{\Gamma (s)}\left(\sum_{\substack{ k=0\\k\neq \dim M }}^{\infty}\frac{a_k(\mathcal{D})}{s+\frac{k-\dim
M}{2}}+\frac{a_{\dim M}(\mathcal{D})-\dim\ker\mathcal{D}}{s}+r(s;D)\right),
\end{equation}
where $r(s;D)$ is an analytic function. The zeta function has a meromorphic extension over $\mathbb{C}$ with simple poles at $s_k=\frac{\dim
M-k}{2}$ for $k\in\mathbb{N}_0$ and residue $Res_{s=s_k}[\zeta (s;\mathcal{D})]=\frac{a_k(\mathcal{D})}{\Gamma (\frac{\dim M-k}{2})}$ at
$s=s_k$. Since $\lim_{s\rightarrow 0}s\Gamma (s)=1$, the zeta function is analytic in the origin and one finds that $\zeta
(0;\mathcal{D})=a_{\dim M}(\mathcal{D})-\dim\ker \mathcal{D}$. In our case this leads to

\begin{equation}
\zeta (0;\Delta _p^{\mathbb{T}_{\beta}^1\times X})\equiv\zeta (0;\Delta _p^{\mathbb{T}_{\beta}^1\times
X}|_{\mathcal{H}^p(\mathbb{T}_{\beta}^1\times X)^{\perp}})=a_{n+1}(\Delta _p^{\mathbb{T}_{\beta}^1\times X})-b_p(\mathbb{T}_{\beta}^1\times
X)\quad\text{for}\ p=0,1.
\end{equation}
Let us denote the eigenvalues of $\Delta _s^{X}$ by $\nu _{l_s}(\Delta _s^{X})$, where $l_s$ runs over an appropriate subset
$J^{(s)}\subset\mathbb{Z}$. Notice that the eigenvalues of $\Delta _r^{\mathbb{T}_{\beta}^1}$ ($r=0,1$) are given by $\nu _{k}(\Delta
_r^{\mathbb{T}_{\beta}^1})=(\frac{2\pi k}{\beta})^2$ with $k\in\mathbb{Z}$. Then the spectrum of $\Delta _p^{\mathbb{T}_{\beta}^1\times
X}|_{\mathcal{H}^p(\mathbb{T}_{\beta}^1\times X)^{\perp}}$ ($p=0, 1$) contains the following set of non-vanishing real numbers

\begin{equation}
\begin{split}
& Spec(\Delta _p^{\mathbb{T}_{\beta}^1\times X}|_{\mathcal{H}^p(\mathbb{T}_{\beta}^1\times X)^{\perp}})= \\ &=\{\nu _{(k,l_s)}^{(r,s)}:=\nu
_{k}(\Delta _r^{\mathbb{T}_{\beta}^1})+\nu _{l_s}(\Delta _s^{X})\neq 0|r+s=p,\ (k,l_s)\in \mathbb{Z}\times J^{(s)}\}.\label{spectrum}
\end{split}
\end{equation}
In terms of these eigenvalues, the corresponding zeta functions can be written as convergent infinite series (for $\Re (s)> \frac{n+1}{2}$)

\begin{equation}
\zeta (s;\Delta _p^{\mathbb{T}_{\beta}^1\times X}|_{\mathcal{H}^p(\mathbb{T}_{\beta}^1\times X)^{\perp}})= \begin{cases} \sum _{(k,l_0)\in
\mathbb{Z}\times J^{(0)}}\left[\nu _{(k,l_0)}^{(0,0)}\right]^{-s}, &\text{if $p=0$},\\ \sum _{(k,l_0)\in \mathbb{Z}\times J^{(0)}}\left[\nu
_{(k,l_0)}^{(1,0)}\right]^{-s}+\sum _{(k,l_1)\in \mathbb{Z}\times J^{(1)}}\left[\nu _{(k,l_1)}^{(0,1)}\right]^{-s}, &\text{if
$p=1$},\end{cases}\label{zeta-general}
\end{equation}
where each eigenvalue appears as often as its multiplicity. Remark that $\nu _{(k,l_0)}^{(0,0)}=\nu _{(k,l_0)}^{(1,0)}$.\par

We notice that there is an intrinsic ambiguity in the definition of the zeta function \eqref{zeta} due to the fact that the eigenvalues of the
Laplace operators are not dimensionless. As a consequence the partition function would admit a dimension as well. To restore this, an
appropriate scale factor $\mu$ of mass dimension $[\mu ]=1$ has to be introduced \cite{hawking}. Instead of \eqref{zeta} one introduces
therefore the dimensionless zeta function $\zeta _{\mu} (s;\mathcal{D}):=\sum _{\nu _{\alpha}(\mathcal{D})\neq 0}(\mu ^{-2}\nu
_{\alpha}(\mathcal{D}))^{-s}$. Formally, this gives $\zeta _{\mu} (s;\mathcal{D})=\mu^{2s}\zeta (s;\mathcal{D})$. In the following all
determinants appearing in \eqref{partition-final} are expressed in terms of these scale dependent but dimensionless zeta functions, i.e.

\begin{equation}
\det _{\mu}{(\Delta _p^{\mathbb{T}_{\beta}^1\times X}|_{\mathcal{H}^p(\mathbb{T}_{\beta}^1\times
X)^{\perp}})}:=\exp{\left[-\frac{d}{ds}|_{s=0}\ \zeta _{\mu} (s;\Delta _p^{\mathbb{T}_{\beta}^1\times
X}|_{\mathcal{H}^p(\mathbb{T}_{\beta}^1\times X)^{\perp}})\right]},\qquad p=0,1.\label{zeta-original}
\end{equation}
However, even with this substitution the partition function \eqref{partition-final} is not dimensionless. This traces back to the occurrence of
the two Jacobian determinants \eqref{group-volume-relation} and \eqref{metric-1}, whose dimensions must be corrected accordingly. In our
convention the coupling constant is set to $1$, which leads to the mass dimensions $[A]=\frac{n-1}{2}$ and $[F_A]=\frac{n+1}{2}$ for the
components of $A$ and $F_A$, respectively. Consequently, one obtains the mass dimensions $[h_X^{(1)}]=-1$ and $[h_X^{(2)}]=1$.\par

In order to get a dimensionless partition function we propose the following replacements in \eqref{group-volume-relation} and \eqref{metric-0},
namely

\begin{equation}
\begin{split}
& \beta V\mapsto\beta V \mu^{n+1}\\
& h_{\mathbb{T}_{\beta}^1\times X}^{(1)}\mapsto diag\left(\beta^{-1} V\mu^{n-1},\beta h_{X}^{(1)}\mu ^{2}\right).
\end{split}\label{dimensionless-jacobian}
\end{equation}
Alternatively, this substitution could be implemented formally by choosing the normalization constant $N=\mu ^{b_1(X)-1}$ in
\eqref{partition-final}.\par

As a result, the free energy $\mathcal{F}^{\theta}(\beta ;V) =-\frac{1}{\beta}\ln{\mathcal{Z}^{\theta}(\beta ;V)}$ admits now the form

\begin{equation}
\begin{split}
\mathcal{F}^{\theta}(\beta ;V) = &\frac{2-b_1(X)}{2\beta}\ln{\beta}-\frac{1}{2\beta}\ln{\det h_X^{(1)}}+\frac{1}{\beta}\zeta ^{\prime}(0;\Delta
_0^{\mathbb{T}_{\beta}^1\times X}|_{\mathcal{H}^0(\mathbb{T}_{\beta}^1\times X)^{\perp}})\\ &-\frac{1}{2\beta}\zeta ^{\prime}(0;\Delta
_1^{\mathbb{T}_{\beta}^1\times X}|_{\mathcal{H}^1(\mathbb{T}_{\beta}^1\times X)^{\perp}})-\frac{1}{\beta}
\ln{\Theta _{b_1(X)}\left(\vec{\theta} |2\pi\sqrt{-1}\ \beta ^{-1}\ h_X^{(1)}\right)}\\
&-\frac{1}{\beta}\ln{\Theta _{b_2(X)}\left(0 |2\pi\sqrt{-1}\ \beta\ h_X^{(2)}\right)}-\frac{b_1(X)}{2\beta}\ln{2\pi}-
\frac{1}{\beta}\ln|TorH^2(X;\mathbb{Z})| \\
&+\biggl(\frac{1}{\beta}\ \zeta (0;\Delta _0^{\mathbb{T}_{\beta}^1\times X}|_{\mathcal{H}^0(\mathbb{T}_{\beta}^1\times
X)^{\perp}})-\frac{1}{2\beta}\ \zeta (0;\Delta _1^{\mathbb{T}_{\beta}^1\times X}|_{\mathcal{H}^1(\mathbb{T}_{\beta}^1\times
X)^{\perp}})+\frac{1-b_1(X)}{2\beta}\biggl)\ln{\mu ^2}.
\end{split}\label{free-energy-generic}
\end{equation}
In the next step we want to express the free energy in terms of the geometry of the spatial manifold $X$: Given any non-negative, self-adjoint
elliptic operator $D$ of second order on $X$ with spectrum $\{\nu_{\alpha}(D)|\alpha\in J\}$, we define the following two auxiliary quantities

\begin{equation}
\begin{split}
\mathcal{I}(s;D):&=\sum _{(k,\alpha )\in (\mathbb{Z}\times J)^{\prime}}\left[ (\frac{2\pi k}{\beta})^2+\nu _{\alpha}(D)\right]^{-s}\\
\hat{\mathcal{I}}(s;D):&=\sum _{k\in\mathbb{Z}}\sum _{\alpha\in J^{\prime}}\left[ (\frac{2\pi k}{\beta})^2+\nu
_{\alpha}(D')\right]^{-s}.\label{zeta-reduction-1}
\end{split}
\end{equation}
The prime in the first formula indicates that the sum runs only over those indices $(k,\alpha )$ such that $(\frac{2\pi k}{\beta})^2+\nu
_{\alpha}(D)\neq 0$. In the second line the index set $J'\subseteqq J$ labels the eigenvalues of the restricted operator $D'$. One immediately
finds that

\begin{equation}
\mathcal{I}(s;D)=\hat{\mathcal{I}}(s;D)+2(\dim\ker D)\ \left(\frac{\beta}{2\pi}\right)^{2s}\zeta _R(2s),\label{zeta-reduction-1-2}
\end{equation}
where $\zeta _R(s)$ is the Riemann zeta function. Hence \eqref{zeta-general} can be rewritten in the form

\begin{equation}
\zeta (s;\Delta _p^{\mathbb{T}_{\beta}^1\times X}|_{\mathcal{H}^p(\mathbb{T}_{\beta}^1\times X)^{\perp}})= \begin{cases} \mathcal{I}(s;\Delta
_0^{X}) &\text{if $p=0$},\\\mathcal{I}(s;\Delta _0^{X})+\mathcal{I}(s;\Delta _1^{X}) \, &\text{if $p=1$}.\end{cases}\label{zeta-general-2}
\end{equation}
Applying the Mellin transform to $\hat{\mathcal{I}}(s;D)$ and performing the integration over $t$ give

\begin{equation}
\begin{split}
\hat{\mathcal{I}}(s;D) &=\frac{1}{\Gamma (s)}\int _0^{\infty}dt\ t^{s-1}\sum _{k\in\mathbb{Z}}e^{-(\frac{2\pi}{\beta})^2k^2t}\sum _{\alpha\in
J^{\prime}}e^{-\nu _{\alpha}(D')t}\\ &= \frac{\beta}{2\sqrt{\pi}\Gamma (s)}\int _0^{\infty}dt\ t^{s-\frac{3}{2}}\sum _{\alpha\in
J^{\prime}}e^{-\nu _{\alpha}(D')t}+\frac{\beta}{\sqrt{\pi}\Gamma (s)}\int _0^{\infty}dt\ t^{s-\frac{3}{2}}\sum
_{k=1}^{\infty}e^{-(\frac{k^2\beta ^2}{4t})}\sum _{\alpha\in J^{\prime}}e^{-\nu _{\alpha}(D')t}\\ &= \frac{\beta}{2\sqrt{\pi}}\frac{\Gamma
(s-{\frac{1}{2}})}{\Gamma (s)}\zeta (s-{\frac{1}{2}};D)+\frac{2^{-s-\frac{3}{2}}\beta ^{s+\frac{1}{2}}}{\sqrt{\pi}\Gamma (s)}\sum
_{k\in\mathbb{Z}}\sum _{\alpha\in J^{\prime}}\left[\frac{k}{\sqrt{\nu _{\alpha}(D')}}\right]^{s-\frac{1}{2}}K_{s-\frac{1}{2}}(k\beta\sqrt{\nu
_{\alpha}(D')}).\label{expansion-zeta-function}
\end{split}
\end{equation}
Here $K_{\nu}(s)$ is the modified Bessel function of the second kind \cite{gradshteyn}. In order to calculate the derivative of
$\mathcal{I}^{\prime}(0;D)$ in $s=0$, we notice that $\zeta (s;D)$ admits a Laurent series expansion in $s=-\frac{1}{2}$, namely

\begin{equation}
\zeta (s-\frac{1}{2};D)=\frac{Res_{s=-\frac{1}{2}}\left[ \zeta (s;D)\right]}{s}+FP_{s=-\frac{1}{2}}\left[\zeta (s;D)\right]+\sum
_{k=1}^{\infty}\tilde{\sigma} _ks^k,
\end{equation}
where $FP$ denotes the finite part of the zeta function in $s=-\frac{1}{2}$. Recall that for an arbitrary meromorphic function $f$, the finite
part in $s_0\in\mathbb{C}$ is defined by

\begin{equation}
FP_{s=s_0}[f]:=\lim _{\epsilon\rightarrow 0_{+}}\frac{1}{2\pi\sqrt{-1}}\oint _{|s-s_0|=\epsilon}\frac{f(s)}{s-s_0}\ ds.
\end{equation}
(see e.g. \cite{finite-part-ref} for the definition and properties). For the zeta function of $D$ this implies
\begin{equation}
FP_{s=s_0}[\zeta (s;D)]=\lim _{\epsilon\rightarrow 0}\ \frac{1}{2}\left[\zeta (-\frac{1}{2}-\epsilon;D)+\zeta (-\frac{1}{2}+\epsilon;D)\right].
\end{equation}
Taking the expansions $\frac{1}{\Gamma (s)}=s+\gamma s^2+\mathcal{O}(s^3)$ and $\Gamma (s-\frac{1}{2})=\Gamma (-\frac{1}{2})(1+\Psi
(-\frac{1}{2})s+\mathcal{O}(s^2))$ for small $s$, where $\gamma$ is the Euler-Mascheroni number and $\Psi$ is the Digamma function
\cite{gradshteyn}, and using that the relation $\lim_{s\rightarrow 0}\frac{d}{ds}(\frac{h(s)}{\Gamma (s)})=h(0)$ holds for any regular function
$h$, one finally obtains

\begin{equation}
\begin{split}
\mathcal{I}^{\prime}(0;D)= & -2(\dim\ker D)\ln\beta -2\sum _{\alpha\in J^{\prime}}\ln \left[ 1-e^{-\beta\sqrt{\nu _{\alpha}(D')}}\right]\\ &
-\beta \left( FP_{s=-\frac{1}{2}}\left[ \zeta (s;D)\right]+2(1-\ln 2)Res_{s=-\frac{1}{2}}\left[ \zeta (s;D)\right]\right).\label{I}
\end{split}
\end{equation}
Using the duality formula \eqref{duality} the heat kernel expansion of $\Delta _p^{\mathbb{T}_{\beta}^1}$ becomes

\begin{equation}
Tr(e^{-t\Delta _p^{\mathbb{T}_{\beta}^1}})=\Theta _1(0|\sqrt{-1}\ \frac{4\pi t}{\beta ^2})=\frac{\beta}{2\sqrt{\pi t}}\Theta
_1(0|\sqrt{-1}\frac{\beta ^2}{4\pi t})\simeq\frac{\beta}{2\sqrt{\pi t}}+\mathcal{O}(e^{-\frac{1}{t}}),\quad \textrm{for}\ t\downarrow 0,
\end{equation}
for $p=0,1$. This implies $a_k(\Delta _p^{\mathbb{T}_{\beta}^1})=\frac{\beta}{2\sqrt{\pi}}\delta_{k,0}$. By comparing the asymptotic expansions
of $\Delta _p^{\mathbb{T}_{\beta}^1\times X}$ and $\Delta _p^{X}$ using \eqref{spectrum}, one finds the following relation between the
corresponding Seeley coefficients,

\begin{equation}
\begin{split}
a_k(\Delta _0^{\mathbb{T}_{\beta}^1\times X}) & = \frac{\beta}{2\sqrt{\pi}}\ a_k(\Delta _0^{X})\\
a_k(\Delta _1^{\mathbb{T}_{\beta}^1\times X}) & = \frac{\beta}{2\sqrt{\pi}} \left[ a_k(\Delta _1^{X})+ a_k(\Delta _0^{X})\right].\label{seeley}
\end{split}
\end{equation}
We want to calculate the sum in the second term of \eqref{I} in the case of $D=\Delta _1^{X}$. Let $\{\nu _{\alpha _{1}}(\Delta _1^{X}|_{im
d_{2}^{\ast}})|\ \alpha_{1}\in J_1\}$ and $\{\nu _{\alpha _{0}}(\Delta _0^{X}|_{im d_{1}^{\ast}})|\ \alpha_{0}\in J_0\}$ be the spectra of
$\Delta _1^{X}|_{im d_{2}^{\ast}}$ and $\Delta _0^{X}|_{im d_{1}^{\ast}}$, respectively. Since the spectrum of $\Delta
_p^{X}|_{\mathcal{H}^p(X)^{\perp}}$ is the union of eigenvalues of $\Delta _p^{X}|_{im d_{p+1}^{\ast}}$ and $\Delta _{p-1}^{X}|_{im
d_{p}^{\ast}}$ the sum runs over $\nu _{\alpha _{1}}(\Delta _1^{X}|_{im d_{2}^{\ast}})$ and $\nu _{\alpha _{0}}(\Delta _0^{X}|_{im
d_{1}^{\ast}})$, respectively. When inserting the explicit expression \eqref{zeta-general-2} into \eqref{free-energy-generic} only the sum over
the eigenvalues of $\Delta _1^{X}|_{im d_{2}^{\ast}}$ survives. Furthermore, this also implies $\zeta (s;\Delta _1^{X}|_{imd_2^{\ast}})= \zeta
(s;\Delta _1^{X})-\zeta (s;\Delta _0^{X})$. In terms of the Seeley coefficients, the residue of $\zeta (s;D)$ in $s=-\frac{1}{2}$ is given by

\begin{equation}
Res_{s=-\frac{1}{2}}\left[\zeta (s;D)\right]=-\frac{a_{\dim X+1}(D)}{2\sqrt{\pi}}.
\end{equation}
Due to the linear structure of $FP$, $Res$ and applying the duality formula \eqref{duality} once again to the $b_1(X)$-dimensional Riemann
Theta function, we finally get the functional integral representation for the free energy

\begin{equation}
\begin{split}
\mathcal{F}^{\theta}(\beta ;V)=& \frac{1}{2}FP_{s=-\frac{1}{2}}\left[\zeta (s;\Delta _1^{X}|_{im
d_{2}^{\ast}})\right]+\frac{1}{2}Res_{s=-\frac{1}{2}}\left[\zeta (s;\Delta _1^{X}|_{im d_{2}^{\ast}})\right]\ln{\left(\frac{e\mu}{2}\right)^{2}}\\
& +\frac{1}{\beta} \sum _{\alpha _1\in J_{1}}\ln\left[ 1-e^{-\beta \sqrt{\nu _{\alpha _1}(\Delta _1^{X}|_{im d_{2}^{\ast}})}}\right]-
\frac{1}{\beta}\ln \Theta
_{b_1(X)}\begin{bmatrix} \vec{\theta} \\ 0\end{bmatrix}\left( 0|\frac{\sqrt{-1}}{2\pi}\beta (h_X^{(1)})^{-1}\right) \\
&-\frac{1}{\beta}\ln{\Theta _{b_2(X)} \left( 0|2\pi\sqrt{-1}\beta (h_X^{(2)})\right)}-\frac{1}{\beta}\ln|TorH^2(X;\mathbb{Z})|.
\end{split}\label{free-energy-final}
\end{equation}
It will be shown in the next section that the Hamiltonian formalism gives exactly the same result. Eq. \eqref{free-energy-final} shows
explicitly that the ambiguity in the free energy is independent of the temperature.\par

Before closing this section we want to annotate briefly why the conventional Faddeev-Popov procedure is not applicable in the present case (For
a detailed account refer to \cite{kelnhofer}). In the Faddev-Popov approach the starting point would be the functional integral
\eqref{partition-total}, however with the choices $\Lambda _{reg}^{\mathcal{G}}=\Lambda _{reg}^{\mathcal{G}_{\ast}}=0$. If the gauge fields
were constrained by the covariant gauge condition $d^{\ast}(A-A_0)=0$, this would lead to the well known additional gauge fixing term
$\frac{1}{2}\|d_1^{\ast}(A-A_0)\|^2$ in the action. This term in combination with the classical Maxwell action would then give the unrestricted
kinetic Laplace operator $\Delta _1^{\mathbb{T}_{\beta}^1\times X}$ acting on the thermal gauge fields in the total action functional. Since
$\mathcal{H}^1(\mathbb{T}_{\beta}^1\times X)\neq 0$ this operator is not invertible, so that the functional integral over the gauge fields
would diverge. This is precisely the Gribov problem in the abelian case. Even if this problem was solved ex post by restricting the Laplace
operator to the non-harmonic forms, the Faddeev-Popov determinant related to the covariant gauge condition would be - following the
conventional approach - only the factor $\det\Delta _0^{\mathbb{T}_{\beta}^1\times X}|_{\mathcal{H}^0(\mathbb{T}_{\beta}^1\times X)}$.\par

In our approach there is an additional multiplicative factor $\det h_{\mathbb{T}_{\beta}^1\times X}^{(1)}$ which turns out to be essential in
the finite temperature context and which appears naturally as part of the Jacobian of the transformation \eqref{diffeo}. It gives rise to the
first term on the right hand side of \eqref{free-energy-generic}. Its first part, namely $\frac{1}{\beta}\ln\beta$, is related to
$\mathcal{H}^1(\mathbb{T}_{\beta}^1)$ and is present even if $b_1(X)=0$. However, this term is compensated by the zero-mode subtraction coming
from the expansion \eqref{zeta-reduction-1-2} of the zeta function of the Laplace operator on $\mathbb{T}_{\beta}^1\times X$ in terms of the
zeta function associated to the corresponding Laplace operators on $X$. Without this cancellation, the entropy \eqref{thermodynamic-funct}
would have admitted a logarithmic divergence in the zero-temperature limit . In this context we would like to refer to the scalar case, where
this logarithmic term was present in \cite{Spraefico} but disappeared in \cite{Lim-Teo} due to a modification of the functional integral
formula for the free energy. By this modification the zero modes of the scalar Laplacian are taken into account correctly and the equality with
the operator (Hamiltonian) approach is provided.\par

The second contribution, namely $\frac{b_1(X)}{2\beta}\ln\beta$ (i.e. the second term in \eqref{free-energy-generic}) is completely absorbed in
the Riemann Theta function due to its duality property. Moreover, our treatment of the Gribov problem automatically rules out the zero modes of
the kinetic Laplace operator right from the beginning, so that the functional integral can be carried out, yet giving a finite result.\par

\section{The free energy in the Hamiltonian approach}

In this section the quantization of Maxwell theory is studied from the canonical (Hamiltonian) point of view. The aim is to determine the
corresponding Hamilton operator $\hat{H}^{\theta}$ in the presence of $\theta$-vacua and to calculate the (thermal) partition function and the
free energy according to

\begin{equation}
\begin{split}
& \hat{\mathcal{Z}}^{\theta}(\beta ,V)=Tr \left( e^{-\beta \hat{H}^{\theta}}\right)\\
& \hat{\mathcal{F}}^{\theta}(\beta ,V)=-\frac{1}{\beta}\ln\hat{\mathcal{Z}}^{\theta}(\beta ;V).\label{partition-hamilton}
\end{split}
\end{equation}
The trace is taken along the physical (i.e. gauge invariant) states. If the theory possesses different topological sectors (like it is in our
case), one has to perform in addition a sum over these sectors as well. We will return to this topic below. In a first step the objects in
\eqref{partition-hamilton} are marked with a caret in order to distinguish them from the partition function and free energy obtained in the
functional integral formalism. \par

As in the previous section, let $Q$ be a principal $U(1)$-bundle over $\mathbb{T}_{\beta}^1\times X$ and consider the pull-back bundle
$\underline{Q}:=\hat{\imath}_{X}^{\ast}Q$ over $X$, where $\hat{\imath}_{X}\colon X\hookrightarrow\mathbb{T}_{\beta}^1\times X$ is the
canonical inclusion $\hat{\imath}_{X}(x):=(1,x)$. For the fixed-time canonical formalism we consider - as usual - the principal $U(1)$-bundle
$\mathbb{R}\times\underline{Q}$ over the space-time manifold $\mathbb{R}\times X$.\par

Let $\mathcal{A}^{\underline{Q}}$ denote the space of connections of $\underline{Q}$ and let $\mathcal{A}^s=\Omega ^0(X;\mathfrak t^1)$ be the
space of scalar gauge potentials. The tangent bundle
$T(\mathcal{A}^{\underline{Q}}\times\mathcal{A}^s)\cong\mathcal{A}^{\underline{Q}}\times\mathcal{A}^s\times\Omega ^1(X;\mathfrak
t^1)\times\Omega ^0(X;\mathfrak t^1)$ is the corresponding configuration space, which is parametrized by the coordinates
$(\underline{A},\underline{A}^s,\underline{\dot{A}},\underline{\dot{A}}^s)$. After performing the analytic continuation, the topological action
$S_{\theta}$ \eqref{theta-action} becomes real. The Lagrangian associated to the classical Maxwell action $S_{tot}$ admits the form

\begin{equation}
L^{\theta}(\underline{A},\underline{A}^s,\underline{\dot{A}},\underline{\dot{A}}^s)=\frac{1}{2}\|\underline{\dot{A}}-d_0\underline{A}^s\|_{\gamma}^2-
\frac{1}{2}\|F_{\underline{A}}\|_{\gamma}^2+\frac{1}{2\pi}\hat{\gamma}(\underline{\dot{A}}-d_0\underline{A}^s,\underline{\theta}),
\end{equation}
where $F_{\underline{A}}=d_1\underline{A}$ is the magnetic field. Here $\|\|_{\gamma}$ refers to the norm which is induced by the metric
$\hat\gamma (\upsilon _1,\upsilon _2):=-\int _{X}\upsilon _1\wedge \star _{\gamma}\upsilon _2$ on $\mathcal{A}^{\underline{Q}}$, where
$\upsilon _1, \upsilon _2\in\Omega ^1(X;\mathfrak t^1)$. Since
$\hat{\gamma}(\underline{\dot{A}}-d\underline{A}^s,\underline{\theta})=\hat{\gamma}(\underline{\dot{A}},\underline{\theta})$ holds, the
topological term is a total time derivative, so that the classical equations of motion remain unchanged despite the addition of this particular
term. By performing the Legendre transformation one gets the corresponding Hamiltonian in the phase space
$T^{\ast}(\mathcal{A}^{\underline{Q}}\times\mathcal{A}^s)$ with conjugate momenta

\begin{equation}
\underline{\Pi}=\frac{\delta
L^{\theta}}{\delta\underline{\dot{A}}}=\underline{\dot{A}}-d\underline{A}^s+\frac{\underline{\theta}}{2\pi},\qquad\underline{\Pi}^s=\frac{\delta
L^{\theta}}{\delta\underline{\dot{A}}^s}=0.
\end{equation}
The phase space $T^{\ast}(\mathcal{A}^{\underline{Q}}\times\mathcal{A}^s)$ is equipped with the canonical symplectic form. Let
$\underline{A}_0\in\mathcal{A}^{\underline{Q}}$ be a fixed background connection satisfying $d_2^{\ast}F_{\underline{A}_0}=0$. The
non-vanishing Poisson brackets for the basic linear phase space functionals $\underline{A}_{u}:=\hat \gamma(\underline{A}-\underline{A}_0,u)$,
$\underline{A}_{u}^s:=\hat \gamma(\underline{A}^s,v)$, $\underline{\Pi}_{u^{\prime}}:=\hat \gamma(\underline{\Pi},u^{\prime})$ and
$\underline{\Pi}_{v^{\prime}}^s:=\hat \gamma(\underline{\Pi}^s,v^{\prime})$ are given by

\begin{equation}
\{\underline{A}_u,\underline{\Pi}_{u^{\prime}}\}=\hat \gamma(u,u^{\prime}),\qquad \{\underline{A}_v^s,\underline{\Pi}_{v^{\prime}}^s\}=\hat
\gamma(v,v^{\prime}),\label{poisson}
\end{equation}
where $u, u^{\prime}\in\Omega ^1(X;\mathfrak t^1)$ and $v, v^{\prime}\in\Omega ^0(X;\mathfrak t^1)$. The Lagrangian $L^{\theta}$ is singular
and leads to two first class constraints in the phase space, namely $\underline{\Pi}^s=0$ and the Gauss law $d^{\ast}\underline{\Pi} =0$. By
choosing the temporal gauge $\underline{A}^s=0$, the first constraint is solved. The dynamical system is restricted to the submanifold
$\mathcal{C}:=\{(\underline{A},\underline{\Pi})\in T^{\ast}\mathcal{A}^{\underline{Q}}|d^{\ast}\underline{\Pi} =0\}$ and governed by the
induced Hamiltonian
\begin{equation}
H^{\theta}(\underline{A},\underline{\Pi})=\frac{1}{2}\|\underline{\Pi}-\frac{\underline{\theta}}{2\pi}\|_{\gamma}^2+
\frac{1}{2}\|F_{\underline{A}}\|_{\gamma}^2.\label{hamilton}
\end{equation}
The corresponding group of gauge symmetries is $\mathcal{G}^{\underline{Q}}=C^{\infty}(X;U(1))$. Let
$\mathcal{G}_{\ast}^{\underline{Q}}:=\mathcal{G}^{\underline{Q}}/U(1)$ be the restricted gauge group, which acts freely on
$\mathcal{A}^{\underline{Q}}$. In fact, the Gauss law constraint is related to the symmetry under the identity component $\mathcal{G}_{\ast
,0}^{\underline{Q}}$ of $\mathcal{G}_{\ast}^{\underline{Q}}$ and can be solved by factoring out this subgroup. This leads to the quotient space
$\mathcal{C}/\mathcal{G}_{\ast ,0}^{\underline{Q}}$. However, since $\mathcal{G}_{\ast}^{\underline{Q}}$ is not connected, i.e. $\pi
_0(\mathcal{G}_{\ast}^{\underline{Q}})\cong\mathbb{Z}^{b_1(X)}$, this is not the true physical phase space $\mathcal{P}$. That space is
obtained by taking out also the large gauge transformations. As a result $\mathcal{P}=(\mathcal{C}/\mathcal{G}_{\ast,0}^{\underline{Q}})/\pi
_0(\mathcal{G}_{\ast}^{\underline{Q}})= \mathcal{C}/\mathcal{G}_{\ast}^{\underline{Q}}$, which is symplectically equivalent to the cotangent
bundle $T^{\ast}\mathcal{M}_{\ast}^{\underline{Q}}$ of the space of gauge orbits
$\mathcal{M}_{\ast}^{\underline{Q}}=\mathcal{A}^{\underline{Q}}/\mathcal{G}_{\ast}^{\underline{Q}}$. At first sight this seems to be not the
final result, since only the restricted gauge group $\mathcal{G}_{\ast}^{\underline{Q}}$ instead of $\mathcal{G}^{\underline{Q}}$ has been
considered so far. But since $\mathcal{A}^{\underline{Q}}/\mathcal{G}^{\underline{Q}}\cong\mathcal{M}_{\ast}^{\underline{Q}}$,
$T^{\ast}\mathcal{M}_{\ast}^{\underline{Q}}$ can be really viewed as the true physical phase space $\mathcal{P}$ of the classical system.\par

In the Hamiltonian approach the underlying geometrical structure is the principal $\mathcal{G}_{\ast}^{\underline{Q}}$-bundle
$\mathcal{A}^{\underline{Q}}\rightarrow\mathcal{M}_{\ast}^{\underline{Q}}$. In order to obtain an explicit parametrization of
$\mathcal{M}_{\ast}^{\underline{Q}}$, we can apply propositions 1 and 2 of section 2 (by replacing the base manifold
$\mathbb{T}_{\beta}^1\times X$ with $X$ in the assumptions). Thus $\mathcal{M}_{\ast}^{\underline{Q}}$ admits the structure of a trivializable
vector bundle over $\mathbb{T}^{b_1(X)}$ with typical fiber $\underline{\mathcal{N}}:=imd_2^{\ast}\otimes\mathfrak t^1$, where
$d_2^{\ast}\colon\Omega ^2(X;\mathfrak t^1)\rightarrow \Omega ^1(X;\mathfrak t^1)$. The trivialization induces a vector bundle isomorphism
$\phi\colon T^{\ast}(\mathbb{T}^{b_1(X)}\times\underline{\mathcal{N}})\rightarrow T^{\ast}\mathcal{M}_{\ast}^{\underline{Q}}$ between the
corresponding cotangent bundles. If $(z_1,\ldots ,z_{b_1(X)};p_1,\ldots ,p_{b_1(X)})$ are coordinates for $T^{\ast}\mathbb{T}^{b_1(X)}$ and
$(\tau ,\Upsilon)$ are those for $T^{\ast}\underline{\mathcal{N}}$, then $\phi$ is given by

\begin{multline}
\phi (z_1,\ldots ,z_{b_1(X)},\tau ;p_1,\ldots ,p_{b_1(X)},\Upsilon )=\\=\left( \left[ \underline{A}_0+2\pi\sqrt{-1}\sum
_{j=1}^{b_1(X)}\sigma_{a_j}(z_j)\rho _{j}^{(1)}+\tau \right],\frac{\sqrt{-1}}{2\pi}\sum _{j,k=1}^{b_1(X)}(h_{X}^{(1)})_{jk}^{-1}p_j\rho
_{k}^{(1)}+\Upsilon \right),
\end{multline}
where the brackets denote the equivalence class in $\mathcal{M}_{\ast}^{\underline{Q}}$ and $s_{a_j}\colon
\underline{V}_{a_j}\subset\mathbb{T}^1\rightarrow\mathbb{R}$ are the local sections of the universal covering
$\mathbb{R}\rightarrow\mathbb{T}^1$ (as defined in Section 2). It can be easily shown that $\phi$ is globally well defined. Notice that we have
parametrized $T^{\ast}\mathcal{M}_{\ast}^{\underline{Q}}$ by pairs $([\underline{A}],\underline{\Pi} )$ satisfying
$d_1^{\ast}\underline{\Pi}=0$. The inverse map reads

\begin{equation}
\begin{split}
\phi ^{-1}([\underline{A}],\underline{\Pi})= & \biggl( e^{\int _{X}(\underline{A}-\underline{A}_0)\wedge\rho _{1}^{(n-1)}},\ldots , e^{\int
_{X}(\underline{A}-\underline{A}_0)\wedge\rho _{b_1(X)}^{(n-1)}},d_2^{\ast}G_2(F_{\underline{A}}-F_{\underline{A}_0}),\\
&\frac{2\pi}{\sqrt{-1}}\sum _{k=1}^{b_1(X)}(h_{X}^{(1)})_{1k}\int _{X}\underline{\Pi}\wedge\rho _{k}^{(n-1)},\ldots ,\frac{2\pi}{\sqrt{-1}}\sum
_{k=1}^{b_1(X)}(h_{X}^{(1)})_{b_1(X)k}\int _{X}\underline{\Pi}\wedge\rho _{k}^{(n-1)},\\
&d_2^{\ast}G_2d_1\underline{\Pi}\biggl).
\end{split}
\end{equation}
The field strength $F_{\underline{A}_0}=d\underline{A}_0$ of the background gauge potential classifies the topologically non-trivial monopole
configurations. With respect to the chosen Betti-basis (see section 2) one gets

\begin{equation}
F_{\underline{A}_0}=2\pi\sqrt{-1}\sum _{k=1}^{b_2(X)}m_{k}\rho _{k}^{(2)},\qquad m_{k}\in\mathbb{Z}.
\end{equation}
In these adapted coordinates the classical Hamiltonian splits into two independent dynamical subsystems with the phases spaces
$T^{\ast}\mathbb{T}^{b_1(X)}$ and $T^{\ast}\underline{\mathcal{N}}$, respectively. The dynamics is governed by the Hamiltonian $\phi
^{\ast}H^{\theta}=H_{harm}^{\theta}+H_{trans}$, where

\begin{multline} H_{harm}^{\theta}(z_1,\ldots ,z_{b_1(X)};p_1,\ldots , p_{b_1(X)})=\\= \frac{1}{2(2\pi)^2}\sum
_{i,j=1}^{b_1(X)}(h_{X}^{(1)})_{ij}^{-1}\ (p_i-2\pi\theta _i)(p_j-2\pi\theta _j)+\frac{(2\pi)^2}{2}\sum _{k,l=1}^{b_2(X)}(h_{X}^{(2)})_{kl}\
m_km_l.\label{hamilton-adapted-new}
\end{multline}
and

\begin{equation}
H_{trans}(\tau;\Upsilon )=\frac{1}{2}\|\Upsilon\|_{\gamma}^2+\frac{1}{2}\ \hat{\gamma}(\tau ,\Delta _1|_{\underline{\mathcal{N}}}\ \tau ).
\end{equation}
In order to quantize these two subsystems we determine the Poisson brackets between the conjugate variables. Let us define the angles
$q_j=\frac{1}{2\pi\sqrt{-1}}\int _{X}(\underline{A}-\underline{A}_0)\wedge\rho _{j}^{(n-1)}$ for $j=1,\ldots ,b_1(X)$, which can be regarded as
coordinates of $\mathbb{R}^{b_1(X)}$ (i.e. of the universal cover of $\mathbb{T}^{b_1(X)}$). By definition $z_j=e^{2\pi\sqrt{-1}q_j}$ and from
\eqref{poisson} we find $\{q_j,p_k\}=\delta _{jk}$. Each gauge transformation $v\in\mathcal{G}^{\underline{Q}}$ of the gauge fields, i.e.
$\underline{A}\mapsto \underline{A}^{v}$, induces a translation $q_j\mapsto q_j+\alpha_j$, where
$\alpha_j=\frac{1}{2\pi\sqrt{-1}}\int_{c_j}v^{\ast}\vartheta ^{U(1)}\in\mathbb{Z}$ are the winding numbers related to $v$ and $j=1,\ldots
,b_1(X)$. The Poisson bracket of the linear phase space functionals $\tau _{u}:=\hat{\gamma}(\tau ,u)$ and $\Upsilon
_{u^{\prime}}:=\hat{\gamma}(\Upsilon ,u^{\prime})$ yields $\{\tau _{u},\Upsilon _{u^{\prime}}\}=\hat{\gamma}(u,d_2^{\ast}G_2d_1u^{\prime})$,
where $u, u^{\prime}\in\Omega ^1(X;\mathfrak t^1)$.\par

In the Schr\"{o}dinger representation the Hilbert space $\mathfrak H $ of physical states splits into the tensor product $\mathfrak H=\mathfrak
H_{harm}\otimes\mathfrak H_{trans}$ of Hilbert spaces of sections of line bundles over $\mathbb{T}^{b_1(X)}$ and over
$\underline{\mathcal{N}}$, respectively. The $\mathcal{G}^{\underline{Q}}$-invariant metric $\hat{\gamma}$ induces a natural connection in the
bundle $\mathcal{A}^{\underline{Q}}\rightarrow\mathcal{M}_{\ast}^{\underline{Q}}$ by declaring the orthogonal complement to the fibers as
horizontal subbundle. The metric restricted to this subbundle finally induces a metric $\underline{\hat{\gamma}}$ on
$\mathcal{M}_{\ast}^{\underline{Q}}$. With respect to the diffeomorphism
$\mathcal{M}_{\ast}^{\underline{Q}}\cong\mathbb{T}^{b_1(X)}\times\underline{\mathcal{N}}$ this metric splits into the direct sum
$\underline{\hat{\gamma}}=h_X^{(1)}\oplus 1_{\underline{\mathcal{N}}}$, where $1_{\underline{\mathcal{N}}}$ is the induced flat metric on
$\underline{\mathcal{N}}$. Let $vol_{\underline{\mathcal{N}}}$ denote the induced volume form, then the (formal) inner product in $\mathfrak H$
is given by

\begin{equation}
<\Psi _1,\Psi _2>_{\mathfrak H}=\int _{\mathbb{T}^{b_1(X)}}vol_{\mathbb{T}^{b_1(X)}}\ (\det h_X^{(1)})^{\frac{1}{2}}\psi _1\bar{\psi} _2\ \int
_{\underline{\mathcal{N}}}\ vol_{\underline{\mathcal{N}}}\ \varphi _1\bar{\varphi} _2,\quad \Psi _i=\psi _i\otimes\varphi _i,\quad i=1,2.
\end{equation}
The Hamilton operator $\hat{H}_{harm}^{\theta}$ is obtained from \eqref{hamilton-adapted-new} by substituting the classical momenta by the
operators $\hat{p}_i=\frac{1}{\sqrt{-1}}\frac{\partial}{\partial q_i}$. This leads to

\begin{equation}
\hat{H}_{harm}^{\theta}= -\frac{1}{2(2\pi)^2}\sum _{i,j=1}^{b_1(X)}(h_{X}^{(1)})_{ij}^{-1}\ \nabla _i\nabla _j+\frac{(2\pi)^2}{2}\sum
_{k,l=1}^{b_2(X)}(h_{X}^{(2)})_{kl}\ m_km_l,\label{hamilton-adapted-operator}
\end{equation}
where $\nabla _i:=\frac{\partial}{\partial q_i}-2\pi\sqrt{-1}\theta _i$ is the covariant derivative. This system can be interpreted as quantum
theory of a classical particle in $b_1(X)$ dimensions coupled to a constant (functional) electric field $\vec{\theta}$ and moving in an
external constant potential determined by the topological sector $\vec{m}\in\mathbb{Z}^{b_2(X)}$. \par

The energy eigenstates of $\hat{H}_{harm}^{\theta}$ are the wave functions $\psi _{\vec{l}}(q_1,\ldots ,q_{b_1(X)})=e^{2\pi\sqrt{-1}\sum
_{i=1}^{b_1(X)}l_iq_i}$ with eigenvalues

\begin{equation}
\varepsilon _{\vec{l},\vec{m}}^{\theta}=\frac{1}{2}\sum _{i,j=1}^{b_1(X)}(h_{X}^{(1)})_{ij}^{-1} (l_i-\theta _i)(l_j-\theta
_j)+\frac{(2\pi)^2}{2}\sum _{\substack{ k,l=1
\\k<l }}^{b_2(X)}(h_{X}^{(2)})_{kl}\ m_km_l.
\end{equation}
In the previous section we introduced the topological action $S_{\theta}$ and argued that this term accounts for the inequivalent quantum
theories caused by the topologically non-trivial configuration space. Now we will justify this argument: Let us introduce the unitary operator
$U_{\theta}$ by

\begin{equation}
\tilde{\psi}(\vec{q}):=(U_{\theta}\psi )(\vec{q})=e^{-2\pi\sqrt{-1}\sum _{j=1}^{b_1(X)}\theta _jq_j}\psi (\vec{q}),\qquad\vec{q}=(q_1,\ldots
,q_{b_1(X)}).
\end{equation}
Since $U_{\theta}\nabla _iU_{\theta}^{-1}=\frac{\partial}{\partial q_i}$, the $\theta$-dependent term in \eqref{hamilton-adapted-operator} can
be removed giving rise to a new Hamilton operator $\hat{H}_{harm}:=U_{\theta}\hat{H}_{harm}^{\theta}U_{\theta}^{-1}$ with eigenstates
$\tilde{\psi}_{\vec{l}}(\vec{q}):=U_{\theta}\psi _{\vec{l}}$. Since $\hat{H}_{harm}^{\theta}$ and $\hat{H}_{harm}$ have the same spectrum, the
corresponding quantum theories are equivalent. However, the wave functions $\tilde{\psi}$ are no longer single-valued, since $\tilde{\psi}
(\vec{q}+\vec{\alpha})=e^{-2\pi\sqrt{-1}\sum _{j=1}^{b_1(X)}\alpha_j\theta _j}\tilde{\psi} (\vec{q})$ for $\vec{\alpha}\in\mathbb{Z}^{b_1(X)}$.
For each fixed $\vec{\theta}\in\mathbb{R}^{b_1(X)}$ these wave functions can be regarded as sections of the line-bundle
$\mathcal{L}^{\theta}:=\mathbb{R}^{b_1(X)}\times _{\vec{\theta}}\mathbb{C}$ over $\mathbb{T}^{b_1(X)}$, which is associated to the universal
covering $\mathbb{R}^{b_1(X)}\rightarrow\mathbb{T}^{b_1(X)}$ via the unitary irreducible representation $\vec{\alpha}\mapsto
e^{2\pi\sqrt{-1}\vec{\alpha}\vec{\theta}}$. Thus different choices for $\vec{\theta}\notin\mathbb{Z}^{b_1(X)}$ lead to inequivalent quantum
theories.\par

Now we are going to quantize the transversal modes: Let $\{\varpi _{\alpha _{1}}|\alpha _{1}\in J_1\}$ denote an orthonormal basis of
eigenforms satisfying $\Delta _1^{X}|_{im d_{2}^{\ast}}\ \varpi _{\alpha _{1}}=\nu _{\alpha _1}(\Delta _1^{X}|_{im d_{2}^{\ast}})\ \varpi
_{\alpha _{1}}$, where each eigenvalue $\nu _{\alpha _1}(\Delta _1^{X}|_{im d_{2}^{\ast}})$ appears as often as its multiplicity. With respect
to the decompositions $\tau =\sum _{\alpha _{1}\in J_1}\tau _{\alpha _{1}}\varpi _{\alpha _{1}}$ and $\Upsilon =\sum _{\alpha _{1}\in
J_1}\Upsilon _{\alpha _{1}}\varpi _{\alpha _{1}}$ one obtains

\begin{equation}
H_{trans}=\sum _{\alpha _{1}\in J_1}\left[\frac{1}{2}|\Upsilon _{\alpha _{1}}|^2+\frac{1}{2}\ \nu _{\alpha _1}(\Delta _1^{X}|_{im
d_{2}^{\ast}})\ |\tau _{\alpha _{1}}|^2\right],\label{trans-hamilton}
\end{equation}
which is nothing but the (well-known) Hamiltonian of an infinite number of harmonic oscillators with frequencies $\sqrt{\nu _{\alpha _1}(\Delta
_1^{X}|_{im d_{2}^{\ast}})}$. The coefficients are $\tau _{\alpha _{1}}=<A-A_0,\varpi _{\alpha _{1}}>$, $\Upsilon _{\alpha _{1}}=<\Upsilon
,\varpi _{\alpha _{1}}>$, having the properties $\bar\tau_{\alpha _{1}}=-\tau _{\alpha _{1}}$ and $\bar\Upsilon _{\alpha _{1}}=-\Upsilon
_{\alpha _{1}}$. The non-vanishing Poisson brackets are $\{\tau _{\alpha _{1}},\bar\Upsilon _{\alpha _{1}^{\prime}}\}=-<\varpi _{\alpha
_{1}},\varpi _{\alpha _{1}^{\prime}}>=\delta _{\alpha _{1},\alpha _{1}^{\prime}}$, which in the quantum theory are replaced by the commutators
$[\tau _{\alpha _{1}},\bar\Upsilon _{\alpha _{1}^{\prime}}]=\sqrt{-1}\ \delta _{\alpha _{1},\alpha _{1}^{\prime}}$. The spectrum of
\eqref{trans-hamilton} is easily obtained in the Fock representation by introducing annihilation and creation operators
\begin{equation}
\begin{split}
& b_{\alpha _{1}}:=\left(4\nu _{\alpha _1}(\Delta _1^{X}|_{im d_{2}^{\ast}})\right)^{-\frac{1}{4}} \left[\sqrt{-1}\ \sqrt{\nu _{\alpha
_1}(\Delta _1^{X}|_{im d_{2}^{\ast}})}\ \tau _{\alpha _{1}}-\Upsilon _{\alpha _{1}} \right]\\ & b_{\alpha _{1}}^{\dag}:=\left(4\nu _{\alpha
_1}(\Delta _1^{X}|_{im d_{2}^{\ast}})\right)^{-\frac{1}{4}} \left[-\sqrt{-1}\ \sqrt{\nu _{\alpha _1}(\Delta _1^{X}|_{im d_{2}^{\ast}})}\
\bar\tau_{\alpha _{1}}-\bar\Upsilon _{\alpha _{1}}\right],
\end{split}
\end{equation}
which have the non-vanishing commutators $\left[ b_{\alpha _{1}},b_{\alpha _{1}^{\prime}}^{\dag}\right]=\delta _{\alpha _{1},\alpha
_{1}^{\prime}}$. The transverse Hamilton operator admits then the following familiar form

\begin{equation}
\hat H_{trans}=\sum _{\alpha _1\in J_1} \sqrt{\nu _{\alpha _1}(\Delta _1^{X}|_{im d_{2}^{\ast}})}\ \left[b_{\alpha _{1}}^{\dag}b_{\alpha
_{1}}+\frac{1}{2}\right],
\end{equation}
which admits the energy eigenvalues $\varepsilon _{k_{\alpha _1}}^{trans}=\sqrt{\nu _{\alpha _1}(\Delta _1^{X}|_{im d_{2}^{\ast}})}\ (k_{\alpha
_1}+\frac{1}{2})$, where $k_{\alpha _1}\in\mathbb{N}_0$.\par

In order to determine the thermal partition function, we have to take the different topological sectors into account. These are labelled by the
first Chern class (see \eqref{chern-class}), namely

\begin{equation}
c_1(\underline{Q})=\hat{\imath}_{X}^{\ast}c_1(Q)=\sum _{j=1}^{b_2(X)}m_{j} \eta _{j}^{(2)}+ \sum _{k=1}^{w}\ y_{k}t_{k}^{(2)}\in
H^2(X;\mathbb{Z}).
\end{equation}
Since $\hat{H}_{harm}^{\theta}$ depends on the free part of $H^2(X;\mathbb{Z})$ only, one obtains

\begin{equation}
\begin{split}
\hat{\mathcal{Z}}^{\theta}(\beta ;V) &=\sum_{c_1(\underline{Q})\in H^2(X;\mathbb{Z})}Tr \left( e^{-\beta \hat{H}^{\theta}}\right)\\
&=\sum_{\vec{m}\in\mathbb{Z}^{b_2(X)}}\ \sum_{\vec{l}\in\mathbb{Z}^{b_1(X)}}\ \prod_{\alpha _1\in J_1}\sum_{k_{\alpha _1}\in\mathbb{N}_0}\
e^{-\beta\varepsilon _{\vec{l},\vec{m}}^{\theta}}\ e^{-\beta\varepsilon _{k_{\alpha _1}}^{trans}}|TorH^2(X;\mathbb{Z})|.
\end{split}
\end{equation}
where the trace has been calculated by summing over all physical eigenstates. The free energy is then given by

\begin{equation}
\begin{split}
\hat{\mathcal{F}}^{\theta}(\beta ;V) &=\frac{1}{2}\sum _{\alpha _1\in J_1}\sqrt{\nu _{\alpha _1}(\Delta _1^{X}|_{im
d_{2}^{\ast}})}+\frac{1}{\beta} \sum _{\alpha _1\in J_{1}}\ln\left[ 1-e^{-\beta \sqrt{\nu _{\alpha _1}(\Delta _1^{X}|_{im
d_{2}^{\ast}})}}\right]\\ &- \frac{1}{\beta}\ln \Theta _{b_1(X)}\begin{bmatrix} \vec{\theta} \\ 0\end{bmatrix}\left(
0|\frac{\sqrt{-1}}{2\pi}\beta (h_X^{(1)})^{-1}\right) -\frac{1}{\beta}\ln{\Theta _{b_2(X)} \left( 0|2\pi\sqrt{-1}\beta h_X^{(2)}\right)}\\
&-\frac{1}{\beta}\ln|TorH^2(X;\mathbb{Z})|.
\end{split}\label{free-energy-final-hamilton}
\end{equation}
Obviously, the first term in \eqref{free-energy-final-hamilton}, which represents the vacuum energy $\varepsilon _{vac}$ of the transverse
modes of the electromagnetic field is infinite and requires a regularization. We choose zeta-function regularization and introduce the
following function in $s\in\mathbb{C}$
\begin{equation}
\tilde{\varepsilon}(s):=\frac{1}{2}\sum _{\alpha _1}\frac{(\nu _{\alpha _1}(\Delta _1^{X}|_{im
d_{2}^{\ast}}))^{\frac{1}{2}}}{(\tilde{\mu}^{-2}\nu _{\alpha _1}(\Delta _1^{X}|_{im
d_{2}^{\ast}}))^{s+\frac{1}{2}}}=\frac{1}{2}\tilde{\mu}^{2s+1}\zeta (s;\Delta _1^{X}|_{im d_{2}^{\ast}}),\label{regularization-1}
\end{equation}
where $\tilde{\mu}$ is a scale parameter with mass dimension $[\tilde{\mu}]=1$. This step is necessary to assign $\tilde{\varepsilon}$ the
correct mass dimension. A natural choice for the vacuum energy could be to take $\varepsilon_{vac}=\tilde{\varepsilon}(-\frac{1}{2})$. However,
this would be reasonable unless $\zeta (s;\Delta _1^{X}|_{imd_2^{\ast}})$ becomes divergent in $s=-\frac{1}{2}$. In order to account for this
case as well, we follow \cite{BVW} and define the regularized vacuum energy of the transverse modes as finite part of $\tilde{\varepsilon}$,
i.e.

\begin{equation}
\varepsilon _{vac}^{reg}:=FP_{s=-\frac{1}{2}}[\tilde{\varepsilon}(s)].\label{regularized-vacuum}
\end{equation}
This regularization scheme amounts to remove the pole from \eqref{regularization-1}. Since
$FP_{s=s_0}[f_1(s)f_2(s)]=f_1(s_0)FP_{s=s_0}[f_2(s)]+f_1^{\prime}(s_0)Res_{s=s_0}[f_2(s)]$ holds for a function $f_1$ being holomorphic in
$s_0$ and $f_2$ being meromorphic with simple pole at the same point $s_0$, one finally gets

\begin{equation}
\varepsilon _{vac}^{reg}=\frac{1}{2}FP_{s=-\frac{1}{2}}[\zeta (s;\Delta _1^{X}|_{im d_{2}^{\ast}})]+\frac{1}{2}Res_{s=-\frac{1}{2}}[\zeta
(s;\Delta _1^{X}|_{im d_{2}^{\ast}})]\ln\tilde{\mu}^{2},
\end{equation}
which represents the finite vacuum free energy of the transverse modes of the Maxwell field. Let us now substitute the first term in
\eqref{free-energy-final-hamilton} by $\varepsilon _{vac}^{reg}$ and choose $\tilde{\mu}=\frac{e\mu}{2}$, then one finds

\begin{equation}
\hat{\mathcal{F}}^{\theta}(\beta ;V)=\mathcal{F}^{\theta}(\beta ;V).
\end{equation}
Hence we have explicitly verified that both quantization schemes are equivalent. Let us stress the fact that keeping the field independent
Jacobian, taking the quotient by the volume of the total gauge group and finally summing over the different topological sectors have been the
main steps in the functional integral scheme to obtain this equality with the Hamiltonian approach. In the remainder of this paper the caret on
the free energy will thus be omitted.\par

The normalization scale $\mu$ expresses the ambiguity of the free energy which occurs whenever $Res_{s=-\frac{1}{2}}\left[\zeta (s;\Delta
_1^{X}|_{im d_{2}^{\ast}})\right]\neq 0$. In that case renormalization issues have to be considered. Alternatively, this could be stated as
follows: If $\zeta (s;\Delta _1^{X}|_{im d_{2}^{\ast}})$ is finite for $s=-\frac{1}{2}$, the free energy is uniquely determined and the scale
dependency must disappear.\par

In the case of massless scalar fields at finite temperature the relation between these two quantization schemes and in particular the question
regarding the zero modes of the kinetic operator have been discussed some time ago in Refs. \cite{BMO, ET, D-2002} and recently in Ref.
\cite{Lim-Teo}.\par

In order to split $\mathcal{F}^{\theta}(\beta ;V)$ into a temperature independent term $\mathcal{F}_{0}^{\theta}(V)$ and a temperature
dependent contribution $\mathcal{F}_{temp}^{\theta}(\beta ;V)$, we use \eqref{theta-original} to rewrite

\begin{multline}
\Theta _{b_1(X)}\begin{bmatrix} \vec{\theta} \\ 0\end{bmatrix}\left( 0|\frac{\sqrt{-1}}{2\pi}\beta (h_X^{(1)})^{-1}\right)=\\
=e^{-\frac{1}{2}\beta\langle\vec{\theta}\rangle ^{\dag}(h_X^{(1)})^{-1}\langle\vec{\theta}\rangle}\ \Theta _{b_1(X)}\left(
\frac{1}{2\pi\sqrt{-1}}\beta (h_X^{(1)})^{-1}\langle\vec{\theta}\rangle |\frac{\sqrt{-1}}{2\pi}\beta (h_X^{(1)})^{-1}\right).
\end{multline}
According to the modular property of the Riemann Theta function \eqref{modular-property}, $\vec{\theta}$ can be replaced by the translated
vector $\langle\vec{\theta}\rangle$ whose $j$-th component is defined by $\langle\theta _j\rangle :=sgn(\theta
_j)\min_{m_j\in\mathbb{Z}}|m_j-\theta _j|$, where $j=1,\ldots ,b_1(X)$. Hence the components of $\langle\vec{\theta}\rangle$ are restricted to
$|\langle\theta _j\rangle|\leq\frac{1}{2}$ for all $j$.

The expression for the free energy of the quantum Maxwell field at finite temperature admits now its final form

\begin{equation}
\begin{split}
\mathcal{F}^{\theta}(\beta ;V)= &\ \mathcal{F}_{0}^{\theta}(V)+\mathcal{F}_{temp}^{\theta}(\beta ;V)\\ = & \
\frac{1}{2}FP_{s=-\frac{1}{2}}\left[\zeta (s;\Delta _1^{X}|_{im d_{2}^{\ast}})\right]+\frac{1}{2}Res_{s=-\frac{1}{2}}\left[\zeta (s;\Delta
_1^{X}|_{im d_{2}^{\ast}})\right]\ln{\left(\frac{e\mu}{2}\right)^{2}}\\&+\frac{1}{2}\sum_{j,k=1}^{b_1(X)}(h_X^{(1)})_{jk}^{-1}\langle\theta
_j\rangle\langle\theta
_k\rangle+\frac{1}{\beta} \sum _{\alpha _1\in J_{1}}\ln\left[ 1-e^{-\beta \sqrt{\nu _{\alpha _1}(\Delta _1^{X}|_{im d_{2}^{\ast}})}}\right]\\
&-\frac{1}{\beta}\ln \Theta _{b_1(X)}\left( \frac{1}{2\pi\sqrt{-1}}\beta
(h_X^{(1)})^{-1}\langle\vec{\theta}\rangle |\frac{\sqrt{-1}}{2\pi}\beta (h_X^{(1)})^{-1}\right) \\
&-\frac{1}{\beta}\ln{\Theta _{b_2(X)} \left( 0|2\pi\sqrt{-1}\beta h_X^{(2)}\right)}-\frac{1}{\beta}\ln|TorH^2(X;\mathbb{Z})|.
\end{split}\label{free-energy-final-form}
\end{equation}
This formula for the regularized (total) free energy of the photon gas confined to a closed manifold $X$ is the main result of the present
paper and - to the best of our knowledge - has not been stated before. It exhibits clearly how the topology of $X$ affects both  the vacuum
energy and the thermodynamic structure of the system. The vacuum energy $\mathcal{F}_{0}^{\theta}$ is the sum of the regularized vacuum energy
of the transverse modes and the ground-state energy of $\hat{H}_{harm}^{\theta}$ (see \eqref{hamilton-adapted-operator}). The latter vanishes
whenever $\vec{\theta}\in\mathbb{Z}^{b_1(X)}$. In any case the Laplace operators appearing in \eqref{free-energy-final-form} are
correspondingly restricted in order to rule out any zero-modes. This is the consequence of the construction of the partition function by using
either a family of local trivializations in the functional integral approach or an appropriate parametrization of the true phase space in the
Hamiltonian scheme.\par

The free energy is unique only if the zeta function converges at $s=-\frac{1}{2}$. On the other hand, if the first and second cohomology group
of $X$ vanish, the Riemann Theta functions as well as the temperature independent $\theta$-vacuum term disappear and the free energy is
completely determined by the transverse modes. Under these conditions the free energy of the quantum Maxwell theory would be indeed a multiple
of the free energy of a massless scalar gas. The $n$-sphere $X=\mathbb{S}^n$ with $n\neq 1,2$ is a typical example for such a configuration. In
so far we gave a proof for the statement argued in \cite{Parthasarathy, Parthasarathy2}.

In the case $n=1$, the transverse modes are absent so that the free energy is exclusively governed by the harmonic component. For
$X=\mathbb{S}^2$ the $\theta$-states are absent, but since $H^2(\mathbb{S}^2;\mathbb{Z})\cong\mathbb{Z}$, the topological sectors contribute
additionally to the thermal excitations.\par

Let us now compare our result with the free energy thermal contributions of a photon gas in flat Euclidean space confined to a very large box
in Euclidean space. In fact, when considering the thermal excitations in the infinite volume limit, the index $\alpha _1$ in the fourth term in
the second equation in \eqref{free-energy-final-form} becomes the continuous $n$-dimensional wave vector $\vec{k}$. Hence all finite size and
topological effects are neglected. Furthermore, the sum is replaced by an integration with respect to the measure
$\frac{d^n\vec{k}}{(2\pi)^n}$. Using \cite{gradshteyn} one gets finally for the thermal part of the free energy density in that limit

\begin{equation}
(n-1)\int_{\mathbb{R}^n}\ \frac{d^n\vec{k}}{(2\pi)^n}\ \ln{\left(1-e^{-\beta |\vec{k}|}\right)}=-(n-1)\pi^{-\frac{n+1}{2}}\ \Gamma
(\frac{n+1}{2})\ \zeta _{R}(n+1)\ \beta ^{-(n+1)},\label{euclidean}
\end{equation}
where the factor $n-1$ is the number of independent degrees of freedom. Since the topology of Euclidean space is trivial the Riemann Theta
functions are absent. Eq. \eqref{euclidean} is the well-known Stefan-Boltzmann term of black-body radiation.\par

Let us emphasize that in our study the photon gas is confined to a closed spatial manifold $X$. Evidently, there is no space "outside" of $X$,
which could contribute  neither to the vacuum energy nor to the thermodynamic excitations of the system. In that sense,
\eqref{free-energy-final-form} represents the intrinsic free energy of the photon gas. This has to be distinguished from configurations where a
partition separates an inside and an outside region in an ambient space. The resulting (Casimir) force exerted by the photon gas on that
partition is then caused by the difference - the so-called \emph{Casimir free energy} - between the free energy in the inside and outside
region. Typical examples are infinite parallel plates \cite{BD}, rectangular cavities \cite{Lim-Teo, GKM} and piston geometries \cite{HJKS,
Lim-Teo1, Kirsten-Fulling}.\par

As was stated above, in our case the vacuum (Casimir) energy depends, in general, on the normalization scale $\mu$. However, if the
electromagnetic field is confined to a compact and connected cavity with smooth perfectly conducting boundary within $\mathbb{R}^{3}$, the
corresponding Casimir energy was shown to be finite \cite{BGH}. This is based on an explicit computation of the heat kernel expansion of the
corresponding Laplace operators and the fact that relevant contributions from the inside and the outside of the cavity cancel. Hence for that
configuration no renormalization is necessary.\par

In the remainder of this section we want to determine the high-temperature limit of the free energy. The starting expression will be
\eqref{free-energy-generic} together with \eqref{zeta-general-2}. Let us define the auxiliary quantity

\begin{equation}
K(s;D^{\prime}):=2\sum _{k=1}^{\infty}\sum _{\alpha\in J^{\prime}}\left[\left(\frac{2\pi
k}{\beta}\right)^{2}+\nu_{\alpha}(D^{\prime})\right]^{-s},
\end{equation}
where the second sum runs over all eigenvalues of $D^{\prime}$, thus excluding the zero modes of $D$. By separating the zero-modes, the
quantity $\mathcal{I}(s;D)$ in \eqref{zeta-reduction-1} can be alternatively rewritten in the form

\begin{equation}
\mathcal{I}(s;D)=\zeta (s;D^{\prime})+2(\dim{\ker D})\ \left(\frac{\beta}{2\pi}\right)^{2s}\zeta
_{R}(2s)+K(s;D^{\prime}).\label{zeta-reduction-2}
\end{equation}
The Mellin transformation of $K(s;D^{\prime})$ reads

\begin{equation}
K(s;D^{\prime})=\frac{2}{\Gamma (s)}\int _0^{\infty}dt\ t^{s-1}\sum _{k=1}^{\infty}e^{-(\frac{2\pi}{\beta})^2k^2t} \sum _{\alpha\in J}e^{-\nu
_{\alpha}(D^{\prime})t} \label{k1}
\end{equation}
for $\Re (s)>\frac{n}{2}$ but can be analytically continued elsewhere \cite{V}. Its asymptotic expansion for $\beta\rightarrow 0$ can be
obtained by substituting the heat kernel expansion \eqref{seeley1} for the restricted operator $D^{\prime}$ with corresponding coefficients,
denoted by $a_m(D^{\prime})$. The integration term by term gives

\begin{equation}
K(s;D^{\prime}) \simeq 2\left(\frac{2\pi}{\beta}\right)^{-2s}\sum
_{m=0}^{\infty}a_m(D^{\prime})\left(\frac{2\pi}{\beta}\right)^{n-m}\frac{\Gamma (s+\frac{m-n}{2})\zeta _{R}(2s+m-n)}{\Gamma (s)}. \label{k2}
\end{equation}
In order to perform the limit $s\rightarrow 0$ of $K(s;D^{\prime})$ and of its derivative respectively, we notice that the function $\Gamma
(s+\frac{m-n}{2})\zeta _{R}(2s+m-n)$ has simple poles at $m=n$ and $m=n+1$. Using that $E_1(s;1)=2\zeta _R(2s)$, where $E_1$ is the Epstein
zeta function \eqref{epstein-zeta} in one dimension, the reflection formula \eqref{reflection} provides the analytic continuation of that
function. If we take the Laurent expansion of $\zeta _R(s)$ at the pole $s=-1$ and the series expansion for $\frac{1}{\Gamma (s)}$ (see above),
a lengthy calculation yields

\begin{equation}
\begin{split}
\frac{1}{2}\frac{d}{ds}|_{s=0}K(s;D^{\prime})= &\sum _{\substack{ m=0\\ m\neq n\\ m\neq n+1 }}^{\infty}a_m(D^{\prime})\ 2^{n-m}\pi
^{-\frac{1}{2}}\beta ^{m-n}\Gamma
(\frac{n+1-m}{2})\zeta _{R}(n+1-m)\\
& +a_{n+1}(D^{\prime})\frac{\beta}{2\pi ^{\frac{1}{2}}}\left(\ln{\left(\frac{\beta}{4\pi}\right)}+\gamma\right)-a_n(D^{\prime})\ln\beta ,
\end{split}\label{K2}
\end{equation}
where (once again) $\gamma$ is the Euler constant. From $a_m(D)=a_m(D^{\prime})+\delta_{m,n}(\dim\ker D)$ and the spectrum of $\Delta _p^X$
($p=0,1$), it follows that

\begin{equation}
a_m(\Delta _1^{X}|_{im d_{2}^{\ast}})=a_m(\Delta _1^{X})-a_k(\Delta _0^{X})+(1-b_1(X))\delta_{m,n}. \label{seeley-relation}
\end{equation}
Using \eqref{k2}, \eqref{K2}, \eqref{seeley-relation}, the explicit form $a_0(\Delta _p^{X})=(4\pi)^{-\frac{n}{2}}\frac{n!}{p!(n-p)!}V$
\cite{Gilkey} and finally applying the duality formula \eqref{duality} to the Riemann Theta function $\Theta _{b_2(X)}(0|...)$, one obtains
from \eqref{free-energy-generic} the high-temperature asymptotic expansion for the free energy in terms of the coefficients $a_m(\Delta
_1^{X}|_{im d_{2}^{\ast}})$, namely

\begin{equation}
\begin{split}
\mathcal{F}^{\theta}(\beta ;V)\simeq & -(n-1)\ \pi^{-\frac{n+1}{2}}\ \Gamma (\frac{n+1}{2})\ \zeta _{R}(n+1)\ \beta ^{-(n+1)}\ V \\
&-\sum _{\substack{ m=1\\ m\neq n\\ m\neq n+1 }}^{\infty}a_m(\Delta _1^{X}|_{im d_{2}^{\ast}})\ 2^{n-m}\pi
^{-\frac{1}{2}}\beta ^{m-n-1}\ \Gamma (\frac{n+1-m}{2})\ \zeta _{R}(n+1-m)\\
&-\frac{1}{2\beta}\left[\zeta^{\prime}(0;\Delta _1^{X}|_{im d_{2}^{\ast}})+\ln{\left(\frac{\det{(2\pi h_X^{(1)})}}{\det{(2\pi h_X^{(2)})}}\
|TorH^2(X;\mathbb{Z})|^2 \right)}\right]\\ &+\frac{1}{2\beta}\left[ 2a_n(\Delta _1^{X}|_{im d_{2}^{\ast}})+b_1(X)+b_2(X)\right]\ln\beta
\\ &+ Res_{s=-\frac{1}{2}}\left[\zeta (s;\Delta _1^{X}|_{im d_{2}^{\ast}})\right]\left(\ln{\left(\frac{\mu\beta}{4\pi}\right)}+\gamma\right)\\
&- \frac{1}{\beta}\ln\Theta _{b_1(X)}\left( \langle\vec{\theta}\rangle |2\pi\sqrt{-1}\beta ^{-1} h_X^{(1)}\right)-\frac{1}{\beta}\ln\Theta
_{b_2(X)}\left( 0|\frac{\sqrt{-1}}{2\pi} \beta ^{-1} (h_X^{(2)})^{-1}\right).
\end{split}\label{K3}
\end{equation}
The first term highlights the familiar Stefan Boltzmann term in $n$ spatial dimensions (see \eqref{euclidean}) and the remaining terms
represent the modifications caused by the topology of $X$. If $\zeta (s;\Delta _1^{X}|_{im d_{2}^{\ast}})$ is regular at $s=-\frac{1}{2}$, the
$\mu$-dependent term vanishes as expected and yields an unambiguous result. A corresponding formula for a gas of massless scalar fields on a
compact manifold with and without boundary has been firstly derived in \cite{DK1} and generalized in \cite{D}. The high temperature limit for
massless scalar fields in different topologies have been discussed by many authors \cite{Spraefico, BMO, ET, D-2002, Lim-Teo} since then.\par

When concerning the Casimir contribution to the free energy (i.e. the Casimir free energy) in the case of material boundaries (e.g. parallel
plates, pistons, rectangular box) all terms, which are of quantum origin like the Stefan-Boltzmann term are cancelled giving rise to the
classical limit at high temperature \cite{GKM, Lim-Teo1, NLS}.\par

Quite recently, the thermal Casimir effect was reconsidered for several types of fields in the static Einstein and closed Friedmann universe
\cite{BKMR, BMMR}. In order to obtain the Casimir free energy, it was proposed to use the renormalization scheme, which is usually applied in
the case of material boundaries, also for the treatment of the topological Casimir effect at finite temperature. In fact, the Casimir free
energy is the difference between the free energy of the topologically non-trivial manifold and the free energy of the tangential
Euclidean/Minkowski space both filled with thermal radiation. Thereby not only the zero-temperature vacuum energy but even the
finite-temperature contributions are renormalized. As a result the Casimir free energy tends to the classical limit at high temperatures, where
the leading term is linear in temperature.
\bigskip

\section{The equation of state}

Once the free energy $\mathcal{F}^{\theta}$ is determined, the main thermodynamic functions can be computed by the following formulae:

\begin{equation}
U^{\theta}(\beta ,V) =\frac{\partial}{\partial\beta}\biggl(\beta\mathcal{F}^{\theta}(\beta ,V)\biggl),\quad S^{\theta}(\beta ,V) =\beta ^2\
\frac{\partial}{\partial\beta}\ \mathcal{F}^{\theta}(\beta ,V),\quad P^{\theta}(\beta ,V) =-\frac{\partial}{\partial V}\
\mathcal{F}^{\theta}(\beta ,V). \label{thermodynamic-funct}
\end{equation}
Here $U^{\theta}$ is the internal energy, $S^{\theta}$ denotes the entropy and $P^{\theta}$ is the pressure of the photon gas. Since the
thermodynamic functions are periodic under translations $\vec{\theta}\mapsto\vec{\theta}+\vec{m}$, for any $\vec{m}\in\mathbb{Z}^{b_1(X)}$, we
can replace $\vec{\theta}$ by $\langle\vec{\theta}\rangle$.\par

Now we want to study the behavior of the free energy under a constant scale transformation $\beta\mapsto\lambda\beta$ and $V\mapsto \lambda
^nV$ with $\lambda\in\mathbb{R}$. Equivalently, this can be regarded as scale transformation of the metric $g\mapsto\lambda ^2g$. A direct
calculation yields

\begin{alignat}{2}
& vol_{X}\mapsto\lambda ^nvol_{X}, &\qquad\quad \nu _{k}^{(i)}(\Delta _r^{\mathbb{T}_{\beta}^1})\mapsto \lambda ^{-2}\nu _{k}^{(i)}(\Delta
_r^{\mathbb{T}_{\beta}^1})
\nonumber\\
& h_{X}^{(1)}\mapsto\lambda ^{n-2}h_{X}^{(1)}, &\qquad\quad  \nu _{l_j}^{(j)}(\Delta _s^{X})\mapsto \lambda ^{-2}\nu _{l_j}^{(j)}(\Delta
_s^{X})\\
& h_{X}^{(2)}\mapsto\lambda ^{n-4}h_{X}^{(2)},\nonumber
\end{alignat}
with $r,s=0,1$. As a consequence, the free energy displays the following transformation behavior

\begin{equation}
\begin{split}
\mathcal{F}^{\theta}(\lambda\beta ,\lambda ^nV) &=\frac{1}{\lambda}\mathcal{F}^{\theta}(\beta ,V)+\frac{\ln\lambda}{\lambda}
Res_{s=-\frac{1}{2}}[\zeta (s;\Delta _1^{X}|_{im d_{2}^{\ast}})]\\
&+\frac{1}{\lambda\beta}\ln{\left(\frac{\Theta _{b_1(X)}\begin{bmatrix} \langle\vec{\theta}\rangle \\ 0\end{bmatrix}\left(
0|\frac{\sqrt{-1}}{2\pi}\beta
(h_X^{(1)})^{-1}\right) \Theta _{b_2(X)}\left(0|2\pi\sqrt{-1}\beta h_{X}^{(2)}\right)}{\Theta _{b_1(X)}\begin{bmatrix} \langle\vec{\theta}\rangle \\
0\end{bmatrix}\left( 0|\frac{\sqrt{-1}}{2\pi}\beta (h_X^{(1)})^{-1}\lambda ^{3-n}\right) \Theta _{b_2(X)}\left(0|2\pi\sqrt{-1}\beta
h_{X}^{(2)}\lambda ^{n-3}\right)}\right)},\label{trafo}
\end{split}
\end{equation}
showing that the free energy does no longer transform homogenously of degree $-1$. In fact, this shows that the free energy is not an extensive
quantity. The violation of the scale invariance is caused on the one hand by the introduction of a scale ambiguity in the regularization of the
vacuum energy and on the other hand by the harmonic modes of the Maxwell field. This anomaly has an implication for the equation of state. In
fact, taking the derivative of \eqref{trafo} by $\lambda$, setting $\lambda =1$ and using \eqref{thermodynamic-funct} yields

\begin{equation}
\mathcal{F}^{\theta}(\beta ,V) =nP^{\theta}(\beta ,V)V-\beta ^{-1}S^{\theta}(\beta ,V)+\Gamma ^{\theta}(\beta ,V),
\end{equation}
with

\begin{equation}
\begin{split}
\Gamma ^{\theta}(\beta ;V) &=Res_{s=-\frac{1}{2}}[\zeta (s;\Delta _1^{X}|_{im
d_{2}^{\ast}})]-\frac{n-3}{2}\sum_{j,k=1}^{b_1(X)}(h_X^{(1)})_{jk}^{-1}\langle\theta _j\rangle\langle\theta
_k\rangle\\
&+\frac{n-3}{\beta}\frac{\partial}{\partial \lambda}|_{\lambda =1}\ln{\left(\frac{\Theta _{b_1(X)}\left( \frac{\lambda}{2\pi\sqrt{-1}}\beta
(h_X^{(1)})^{-1}\langle\vec{\theta}\rangle |\lambda \frac{\sqrt{-1}}{2\pi}\beta (h_X^{(1)})^{-1}\right)}{\Theta
_{b_2(X)}\left(0|2\lambda\pi\sqrt{-1}\beta h_{X}^{(2)}\right)}\right)}.\label{trafo-1}
\end{split}
\end{equation}
Together with the general relation $\mathcal{F}^{\theta}(\beta ,V)=U^{\theta}(\beta ,V)-\beta ^{-1}S^{\theta}(\beta ,V)$ one gets the following
modified equation of state

\begin{equation}
P^{\theta}(\beta ,V)=\frac{1}{nV}\left[ U^{\theta}(\beta ,V)-\Gamma ^{\theta}(\beta ,V)\right].\label{eq-state}
\end{equation}
Thus we have shown that the equation of state of a photon gas on a topologically non-trivial and compact spatial manifold differs from the
conventional one in the large volume limit by the anomalous term $\Gamma ^{\theta}$. Irrespective of the topology of $X$, the "topological"
terms in $\Gamma ^{\theta}$ vanishes for $n=3$. If in addition the zeta function is finite for $s=-\frac{1}{2}$, then we obtain the familiar
relation $P^{\theta}V=\frac{1}{3}U^{\theta}$ in three dimensions.

\section{Explicit results for the $n$-torus $\mathbb{T}^n$}

In the remainder of this paper we want to consider the photon gas confined to a $n$-dimensional torus $X=\mathbb{T}^n$ in more detail. The
$n$-torus is equipped with a flat metric $\gamma =\sum _{i=1}^{n}L_i^2dt^{i}\otimes dt^{i}$, where the sequence $(t^{i})_{i=1}^n$ denotes the
local coordinates of $\mathbb{T}^n$ and $L_i$ is the length in the $i$-th direction, so that the spatial volume is $V=\prod _{i=1}^{n}L_i$.
This configuration can be equivalently realized as field theory in a $n$-dimensional rectangular box $X=[0,L_1]\times\ldots\times [0,L_n]$
subjected to periodic spatial boundary conditions.\par

Massless scalar fields in a box with periodic boundary conditions at finite temperature were studied in \cite{Amb-Wolf, Spraefico, edery,
Lim-Teo} based on Epstein zeta function regularization and in \cite{edery} using a multidimensional cut-off. An analysis of the finite-size
effects in a universe with toroidal topology was presented in \cite{ZK1}.
\par

We would like to stress once again that in the case of a non-vanishing boundary (where the fields in the rectangular box are satisfying
Dirichlet or von Neumann boundary conditions) one has to take into account the contribution both from the inside of the cavity as well as from
the outside region \cite{GKM} for the calculation of the Casimir free energy.\par

A Betti basis for $\mathcal{H}_{\mathbb{Z}}^1(\mathbb{T}^n)$ is given by $\rho
_{j}^{(1)}=\frac{1}{L_j}\hat{i}_{j}^{\ast}vol_{\mathbb{T}^n}^{\gamma}=dt^j$, where $\hat i_j\colon\mathbb{T}^1\hookrightarrow\mathbb{T}^n$ is
the canonical inclusion of the $j$-th position and $j=1,\ldots ,n$. Correspondingly, the dual Betti basis for
$\mathcal{H}_{\mathbb{Z}}^{n-1}(\mathbb{T}^n)$ is provided by $\rho _{j}^{(n-1)}=(-1)^{j-1}\rho _1^{(1)}\wedge\ldots ,\wedge\widehat{\rho _j
^{(1)}}\wedge\ldots ,\wedge \rho _n ^{(1)}$, where the caret denotes omission of the respective element. By construction $\int
_{\mathbb{T}^n}\rho _j^{(1)}\wedge\rho _k^{(n-1)}=\delta _{jk}$. Furthermore $\rho
_{jk}^{(2)}=\frac{1}{L_jL_k}\hat{i}_{j}^{\ast}vol_{\mathbb{T}^n}^{\gamma}\wedge\hat{i}_{k}^{\ast}vol_{\mathbb{T}^n}^{\gamma}= dt^j\wedge dt^k$
induces a Betti basis for $\mathcal{H}_{\mathbb{Z}}^{2}(\mathbb{T}^n)$, where $j,k=1,\ldots ,n$ and $j<k$. As a consequence,
$\mathcal{H}^{p}(\mathbb{T}^n)$ ($p=1,2$) admit the following metrics

\begin{equation}
\begin{split}
& h_{\mathbb{T}^n}^{(1)}=diag\left(\frac{V}{L_1^2},\ldots ,\frac{V}{L_n^2}\right)\\
& h_{\mathbb{T}^n}^{(2)}=diag\left(\frac{V}{L_1^2L_2^2},\frac{V}{L_1^2L_3^2}\ldots ,\frac{V}{L_{n-1}^2L_n^2}\right).
\end{split}
\end{equation}
In order to obtain the free energy we have to calculate the spectrum and the zeta-function of $\Delta _1^{\mathbb{T}^n}|_{im d_{2}^{\ast}}$. It
is easy to prove that the eigenvalues of $\Delta _1^{\mathbb{T}^n}|_{im d_{2}^{\ast}}$ are the sequence of real numbers $\nu _{\vec{m}}(\Delta
_1^{\mathbb{T}^n}|_{im d_{2}^{\ast}})=\sum _{j=1}^n\left(\frac{2\pi\ m_j}{L_j}\right)^2$, parametrized by
$\vec{m}\in\mathbb{Z}_0^n:=\mathbb{Z}^n\backslash 0$. For each $\vec{m}\in\mathbb{Z}_0^n$, we introduce vielbeins $\epsilon _{rj,\vec{m}}$,
which are labelled by $r$ and have components indexed by $j$, satisfying

\begin{equation}
\sum _{j=1}^n\epsilon_{rj,\vec{m}}\ \epsilon _{r^{\prime}j,\vec{m}}=\delta _{rr^{\prime}},\qquad \epsilon _{nj,\vec{m}}=\left(\nu
_{\vec{m}}(\Delta _1^{\mathbb{T}^n}|_{im d_{2}^{\ast}})\right)^{-\frac{1}{2}}\ \frac{2\pi m_j}{L_j},\quad j,r=1,\ldots ,n.
\end{equation}
Let us define the inner product $(\varphi,\chi )_{\gamma}:=<\varphi ,\bar{\chi}>_{\gamma}$ for $\varphi,\chi\in\Omega
^{p}(\mathbb{T}^n,\mathbb{C})$, where $\bar{\chi}$ is the complex conjugate of $\chi$, then the following $1$-forms

\begin{equation}
\varpi _{r,\vec{m}}:=\sum _{j=1}^n \epsilon _{rj,\vec{m}}\ \left(\frac{L_j^2}{V}\right)^{\frac{1}{2}}\ e^{2\pi\sqrt{-1}\sum _{i=1}^nm_it_i}\
dt^{j},\qquad \vec{m}\in\mathbb{Z}_0^n,\ r=1,\ldots ,n
\end{equation}
satisfy $(\varpi _{r,\vec{m}},\varpi _{r^{\prime},\vec{m}^{\prime}})_{\gamma}=\delta _{rr^{\prime}}\delta _{\vec{m},\vec{m}^{\prime}}$. A
direct calculation gives

\begin{equation}
d_2^{\ast}G_2d_1\varpi _{r,\vec{m}}=
\begin{cases}
\varpi _{r,\vec{m}} & \text{if } r=1,\ldots ,n-1 \\
0 & \text{if } r=n,
\end{cases}
\end{equation}
and

\begin{equation}
\Delta _1^{\mathbb{T}^n}|_{im d_{2}^{\ast}}\ \varpi _{r,\vec{m}}=\nu _{\vec{m}}(\Delta _1^{\mathbb{T}^n}|_{im d_{2}^{\ast}})\ \varpi
_{r,\vec{m}},\qquad \vec{m}\in\mathbb{Z}_0^n,\ r=1,\ldots , n-1,\label{eigenvalue}
\end{equation}
proving that the set $\{\varpi _{r,\vec{m}}|r=1,\ldots ,n-1,\ \vec{m}\in\mathbb{Z}_0^n\}$ provides an orthonormal basis of eigenforms of
$\Delta _1^{\mathbb{T}^n}|_{im d_{2}^{\ast}}$ with respect to the inner product $(,)_{\gamma}$. Evidently, the multiplicity of each eigenvalue
is just $(n-1)$. This is precisely the number of independent polarization states of the electromagnetic field. In terms of the Epstein zeta
function \eqref{epstein-zeta} we see that

\begin{equation}
\zeta (s;\Delta _1^{\mathbb{T}^n}|_{im d_{2}^{\ast}})=(n-1)\ E_n(s;\frac{2\pi}{L_1},\ldots ,\frac{2\pi}{L_n}).\label{zeta-torus1}
\end{equation}
The regularized vacuum energy for the transverse modes \eqref{regularized-vacuum} is then given by

\begin{equation}
\varepsilon _{vac}^{reg}:=\frac{1}{2}\lim_{\eta\rightarrow 0}\left[ \left(\frac{e\mu}{2}\right)^{-2\eta}E_n(-\frac{1}{2}-\eta
;\frac{2\pi}{L_1},\ldots ,\frac{2\pi}{L_n})+\left(\frac{e\mu}{2}\right)^{2\eta}E_n(-\frac{1}{2}+\eta ;\frac{2\pi}{L_1},\ldots
,\frac{2\pi}{L_n})\right].
\end{equation}
Applying the reflection formula \eqref{reflection} and performing the limit $\eta\rightarrow 0$ leads to

\begin{equation}
\varepsilon _{vac}^{reg}=-\frac{n-1}{2}\left[\prod _{j=1}^nL_j\right]\ \pi ^{-\frac{n+1}{2}}\Gamma(\frac{n+1}{2})E_n(\frac{n+1}{2};L_1,\ldots
,L_n).\label{vacuum-energy}
\end{equation}
By construction the vacuum free energy is finite which is - of course - confirmed by the fact that $E_n(\frac{n+1}{2};L_1,\ldots , L_n)$
converges and is positive for every $n$ \cite{LT-zeta}. Moreover, for any given $s$, the minima of $E_n(s;L_1,\ldots , L_n)$ with fixed volume
$V=\prod_{i=1}^{n}$ appears at $L_1=\cdots =L_n$.\par

As a consequence, there is no ambiguity in defining a finite vacuum energy. Inserting this result into \eqref{free-energy-final-form}, one
immediately obtains the expression for the free energy on the $n$-torus, namely

\begin{equation}
\begin{split}
\mathcal{F}^{\theta}(\beta ;V) &=-\frac{n-1}{2}\ V\ \pi ^{-\frac{n+1}{2}}\Gamma(\frac{n+1}{2})E_n(\frac{n+1}{2};L_1,\ldots
,L_n)+\frac{1}{2V}\sum_{j=1}^{n}L_j^2\langle\theta _j\rangle^2\\ &+ \frac{n-1}{\beta}\sum _{(k_1,\ldots ,k_n)\in\mathbb{Z}_{0}^{n}}\ln{\left[
1-e^{-2\pi\beta\sqrt{\sum _{l=1}^n(\frac{k_l}{L_l})^2}}\right]}
\\ & -\frac{1}{\beta}\sum _{j=1}^{n}\ln{\Theta _1\left( \frac{1}{2\pi\sqrt{-1}}\beta\frac{L_j^2}{V}
\langle\theta _j\rangle|\frac{\sqrt{-1}}{2\pi}\beta \frac{L_j^2}{V}\right)} -\frac{1}{\beta}\sum _{\substack{ j,k=1 \\j<k }}^n\ln{\Theta _1
\left( 0|2\pi\sqrt{-1}\beta\frac{V}{L_j^2L_k^2}\right)}.
\end{split}\label{hamilton_low}
\end{equation}
Notice that $Tor H^2(\mathbb{T}^n;\mathbb{Z})=0$. For $n=1$ we recover the free energy of pure electrodynamics on the circle (see e.g.
\cite{AJW}).
\par

For sake of completeness we want to sketch an alternative calculation of the free-energy which derives directly from
\eqref{free-energy-generic}. The eigenvalues of $\Delta
_p^{\mathbb{T}_{\beta}^1\times\mathbb{T}^n}|_{\mathcal{H}^p(\mathbb{T}_{\beta}^1\times\mathbb{T}^n)^{\perp}}$ are given by

\begin{equation}
\begin{split}
&\nu _{\underline{k}}^{(0,0)}=\left(\frac{2\pi k_0}{\beta}\right)^2+\sum _{i=1}^n\left(\frac{2\pi k_i}{L_i}\right)^2,\quad p=0\\ &\nu
_{\underline{k}}^{(1,0)}=\nu _{\underline{k}}^{(0,1)} =\left(\frac{2\pi k_0}{\beta}\right)^2+\sum _{i=1}^n\left(\frac{2\pi
k_i}{L_i}\right)^2,\quad p=1
\end{split}
\end{equation}
where $\underline{k}=(k_0,k_1,\ldots ,k_n)\in\mathbb{Z}_{0}^{n+1}:=\mathbb{Z}^{n+1}\backslash 0$. Given the multiplicities of the eigenvalues,
the zeta functions of the Laplace operators become proportional to the Epstein zeta functions in $n+1$ dimensions, namely

\begin{equation}
\zeta (s;\Delta _p^{\mathbb{T}_{\beta}^1\times \mathbb{T}^n}|_{\mathcal{H}^p(\mathbb{T}_{\beta}^1\times \mathbb{T}^n)^{\perp}})=\begin{cases}
(n+1)\ E_{n+1}(s;\frac{2\pi}{\beta},\frac{2\pi}{L_1},\ldots ,\frac{2\pi}{L_n}) & \text{if } p=1
\\E_{n+1}(s;\frac{2\pi}{\beta},\frac{2\pi}{L_1},\ldots ,\frac{2\pi}{L_n})& \text{if } p=0.\end{cases} \label{zeta-torus}
\end{equation}
By applying the duality relation \eqref{duality} to $\Theta _1(\theta _j|\beta ^{-1}..)$ in \eqref{free-energy-generic} and using the
Chowla-Selberg formula \eqref{epstein-expansion} by setting $m=n+1$, $l=1$ and $c_1=\frac{2\pi}{\beta}$, $c_2=\frac{2\pi}{L_1}$,...,
$c_{n+1}=\frac{2\pi}{L_n}$ one immediately obtains \eqref{hamilton_low}.\par

In the torus case the vacuum energy is unique and the equation of state is modified only by the Riemann Theta functions. In contrast the free
energy of the scalar massless gas transforms properly under scale transformations \cite{Lim-Teo}.

In the next two subsections we will derive explicit formulae for the thermodynamic functions in the low- and high temperature regimes,
respectively. In order to simplify the calculations we will consider hereafter a \textit{uniform} $n$-torus, i.e. $L_1=\cdots
=L_n=V^{\frac{1}{n}}$.

\subsection{The low temperature regime}

It is evident that expression \eqref{hamilton_low} is suitable for studying the low temperature behavior ($\beta\gg 1$) of the system. Let us
indicate this fact by a subscript and write $\mathcal{F}_{low}^{\theta}(\beta ;V)$ when the free energy is displayed in the form
\eqref{hamilton_low}. \par

We can give an explicit series expansion for the last two terms in \eqref{hamilton_low}, denoted by $f_{low}^{\theta}$ for further reference.
Since $|\langle\theta _j\rangle|\leq\frac{1}{2}$, the constraint \eqref{constraint} is satisfied. Eq. \eqref{asymptotic-expansion} can be
applied to $f_{low}^{\theta}$ giving the following series expansion

\begin{equation}
\begin{split}
f_{low}^{\theta}(\beta ;V) :=&-\frac{1}{\beta}\ln 2^{\Lambda}- \frac{1}{\beta}\sum_{m=1}^{\infty}\frac{1}{m}\
\frac{n+2(-1)^m\Lambda}{1-e^{\beta V^{\frac{2-n}{n}}m}}\\
& -\frac{1}{\beta}\sum_{\substack{ j=1 \\|\langle\theta _j\rangle|\neq\frac{1}{2} }}^{n}\sum_{m=1}^{\infty}\frac{(-1)^{m+1}}{m}\
\frac{e^{-\frac{\beta }{2} V^{\frac{2-n}{n}}m (1-2\langle\theta _j\rangle)}+e^{-\frac{\beta }{2} V^{\frac{2-n}{n}}m (1+2\langle\theta
_j\rangle)}}{1-e^{-\beta
V^{\frac{2-n}{n}}m}}\\
& -\frac{n(n-1)}{2\beta}\sum_{m=1}^{\infty}\frac{1}{m}\ \frac{1+2(-1)^m e^{\frac{(2\pi)^2}{2}\beta V^{\frac{n-4}{n}}m}}{1-e^{(2\pi)^2\beta
V^{\frac{n-4}{n}}m}},\label{asym-expansion-2}
\end{split}
\end{equation}
where $\Lambda$ is the number of components $\theta _j$ which satisfy $|\langle\theta _j\rangle|=\frac{1}{2}$. Clearly $0\leq\Lambda\leq n$.
Apart from the first term, all others terms decrease exponentially for $\beta\rightarrow\infty$. \par

The vacuum energy $\mathcal{F}_{0}^{\theta}(V)$ is determined by taking the zero temperature limit of \eqref{hamilton_low}

\begin{multline}
\mathcal{F}_{0}^{\theta}(V):=\lim _{\beta\rightarrow\infty}\mathcal{F}_{low}^{\theta}(\beta ;V)=\\ =-\frac{n-1}{2}\ \pi^{-\frac{n+1}{2}}\
\Gamma (\frac{n+1}{2})E_{n}(\frac{n+1}{2};1,\ldots ,1)\ V^{-\frac{1}{n}}+V^{\frac{2-n}{n}}\sum_{j=1}^{n}\frac{\langle\theta _j\rangle^2}{2}.
\end{multline}
If $\vec{\theta}\in\mathbb{Z}^{n}$, the free energy of the Maxwell field equals the sum of the free energy of $(n-1)$ massless scalar fields
(see e.g. the discussion of the massless scalar field with periodic boundary conditions in \cite{Spraefico} and \cite{Lim-Teo}). It is evident,
that $\mathcal{F}_{0}^{\theta}(\lambda ^{n}V)=\lambda ^{-1}\mathcal{F}_{0}^{\theta}(V)$ for $\vec{\theta}\in\mathbb{Z}^{n}$.\par

The internal energy derives from \eqref{hamilton_low} and reads for the uniform $n$-torus

\begin{equation}
\begin{split}
U_{low}^{\theta}(\beta ;V) = &\frac{(1-n)}{2}\ V^{-\frac{1}{n}}\pi^{-\frac{n+1}{2}}\ \Gamma (\frac{n+1}{2})E_{n}(\frac{n+1}{2};1,\ldots
,1)+V^{\frac{2-n}{n}}\sum_{j=1}^{n}\frac{\langle\theta _j\rangle^2}{2}\\ &+ \sum _{\vec{k}\in\mathbb{Z}_{0}^{n}}\frac{2\pi
(n-1)V^{-\frac{1}{n}} |\vec{k}|}{e^{2\pi\beta V^{-\frac{1}{n}}|\vec{k}|}-1}+\ \frac{\partial}{\partial\beta}\left(\beta f_{low}^{\theta}(\beta
;V)\right),
\end{split}\label{energy-tot-low-V}
\end{equation}
where we have defined the abbreviations $\vec{k}=(k_1,\ldots ,k_n)$ and $|\vec{k}|=(\sum _{l=1}^nk_l^2)^{1/2}$. A direct calculation using
\eqref{asym-expansion-2} shows that the contributions in the last term of \eqref{energy-tot-low-V} decrease exponentially for low temperatures.
One obtains

\begin{equation}
U_{0}^{\theta}(V):=\lim _{\beta\rightarrow \infty}U_{low}^{\theta}(\beta ;V)=\lim _{\beta\rightarrow\infty}\mathcal{F}_{low}^{\theta}(\beta
;V)=\mathcal{F}_{0}^{\theta}(V).\label{zero-temp-internal}
\end{equation}
The entropy of the system is given by

\begin{equation}
\begin{split}
S_{low}^{\theta}(\beta ;V)= &(1-n)\sum _{\vec{k}\in\mathbb{Z}_{0}^{n}}\ln{\left[ 1-e^{-2\pi\beta
V^{-\frac{1}{n}}|\vec{k}|}\right]}+(n-1)V^{-\frac{1}{n}}\beta\sum _{\vec{k}\in\mathbb{Z}_{0}^{n}}\frac{2\pi |\vec{k}|}{e^{2\pi\beta
V^{-\frac{1}{n}}|\vec{k}|}-1}\\ &+\beta^{2}\ \frac{\partial}{\partial\beta}f_{low}^{\theta}(\beta ;V).
\end{split}\label{entropy-tot-low-V}
\end{equation}
Using the series expansion \eqref{asym-expansion-2} and performing the zero temperature limit only the first term in \eqref{asym-expansion-2}
survives, leading to

\begin{equation}
\lim _{\beta\rightarrow \infty}S_{low}^{\theta}(\beta ;V)=\ln 2^{\Lambda}.\label{entropy-zero}
\end{equation}
This is a consequence of the fact that the ground-state of the system is degenerate of degree $2^{\Lambda}$. To verify this we notice that for
a fixed $\vec{\theta}$ the lowest energy level of $\hat{H}_{harm}^{\theta}$ (see \eqref{hamilton-adapted-operator}) in a given topological
(monopole) sector reads

\begin{equation}
\varepsilon _{0,\vec{m}}^{\theta}=\frac{1}{2}\ V^{\frac{2-n}{n}}\sum _{j=1}^{n} \langle\theta _j\rangle ^2+\frac{(2\pi)^2}{2}\
V^{\frac{n-4}{n}}\sum _{k=1}^{\frac{n(n-1)}{2}}m_k^2.
\end{equation}
Here $\vec{m}=(m_1,\ldots , m_k,\ldots )\in\mathbb{Z}^{\frac{n(n-1)}{2}}$ labels the topological sectors. The degeneracy appears exactly for
the parameter values $\theta _j=\frac{2k_{j}+1}{2}$, $k_{j}\in\mathbb{Z}$ (i.e. $|\langle\theta _j\rangle |=\frac{1}{2}$). Hence we have
explicitly proved that the 3rd law of thermodynamics (Nernst theorem) holds for the photon gas on the $n$-torus. \par

The pressure of the photon gas admits the following form:

\begin{equation}
\begin{split}
P_{low}^{\theta}(\beta ;V)=&\frac{V^{-\frac{n+1}{n}}}{n}\bigg[\sum _{\vec{k}\in\mathbb{Z}_{0}^{n}}\frac{2\pi (n-1) |\vec{k}|}{e^{2\pi\beta
V^{-\frac{1}{n}}|\vec{k}|}-1}- \frac{n-1}{2}\ \pi^{-\frac{n+1}{2}}\ \Gamma (\frac{n+1}{2})E_{n}(\frac{n+1}{2};1,\ldots ,1)\bigg]
\\ &+\frac{n-2}{n}\ V^{\frac{2(1-n)}{n}}\sum_{j=1}^{n}\frac{\langle\theta _j\rangle^2}{2}-\frac{\partial}{\partial V}
f_{low}^{\theta}(\beta ;V).
\end{split}\label{pressure-tot-low-V}
\end{equation}
In the zero temperature limit, the pressure converges to

\begin{multline}
P_{0}^{\theta}(V):=\lim _{\beta\rightarrow \infty}P_{low}^{\theta}(\beta ;V)=\\ =-\frac{n-1}{2n}\ V^{-\frac{n+1}{n}}\pi^{-\frac{n+1}{2}}\
\Gamma (\frac{n+1}{2})E_{n}(\frac{n+1}{2};1,\ldots ,1)+\frac{n-2}{n}V^{\frac{2(1-n)}{n}}\sum_{j=1}^{n}\frac{\langle\theta
_j\rangle^2}{2}.\label{zero-temp-pressure}
\end{multline}
For $n\neq 1$, one gets for the large volume limit $\lim_{V\rightarrow\infty}P_{0}^{\theta}(V)=0$. Additionally, the following four cases can
be easily distinguished:

\begin{description}
    \item[a) $n=1$:] $P_{0}^{\theta}(V)=-\frac{\langle\theta\rangle^2}{2}$ is non vanishing whenever $\theta\notin\mathbb{Z}$.
    \item[b) $n=2$:] The pressure is exclusively determined by the (vacuum) transverse modes of the gauge fields. Hence $P_{0}^{\theta}(V)<0$
    for all $V$.
    \item[c) $n=3$:] For $\vec{\theta}\in\mathbb{Z}^3$ one gets $P_{0}^{\theta}(V)<0$ for all $V$.
    \item[d) $n>3$:] For each $\vec{\theta}\notin\mathbb{Z}^n$ there exists a critical volume $V_{crit}$ such that $P_{0}^{\theta}(V_{crit})=0$. For
    $V<V_{crit}$ one has $P_{0}^{\theta}(V)>0$ whereas for $V>V_{crit}$ one has $P_{0}^{\theta}(V)<0$.
\end{description}
A direct comparison of \eqref{zero-temp-internal} and \eqref{zero-temp-pressure} shows that

\begin{equation}
P_{0}^{\theta}(V)=\frac{1}{nV}U_{0}^{\theta}(V)+\frac{n-3}{n}\ V^{\frac{2(1-n)}{n}}\ \sum_{j=1}^{n}\frac{\langle\theta
_j\rangle^2}{2}.\label{casimir-pressure}
\end{equation}
Alternatively this result may be obtained as the zero-temperature limit of the equation of state \eqref{eq-state}.

\subsection{The high temperature regime}

An appropriate expression - hereafter called $\mathcal{F}_{high}^{\theta}(\beta ;V)$ -  for analyzing the high-temperature regime ($\beta\ll
1$) can be obtained from \eqref{free-energy-generic} and \eqref{zeta-torus} as follows: Taking $m=n+1$, $l=n$, $c_1=\frac{2\pi}{L_1}$,...,
$c_{n}=\frac{2\pi}{L_n}$ and $c_{n+1}=\frac{2\pi}{\beta}$ in \eqref{epstein-expansion} and applying \eqref{duality} one obtains after a lengthy
calculation

\begin{equation}
\begin{split}
\mathcal{F}_{high}^{\theta}(\beta ;V)= &(1-n)\pi^{-\frac{n+1}{2}} \Gamma (\frac{n+1}{2})\zeta _{R}(n+1)\beta ^{-(n+1)}V+
\frac{\kappa_1}{\beta}\ln{\beta}+\frac{1}{\beta}\left(\kappa_2\ln{V}+\kappa_3\ln{2\pi}\right)\\ &-2(n-1)\beta ^{-\frac{n+2}{2}}V \sum
_{k_{n+1}=1}^{\infty}\ \sum _{\vec{k}\in\mathbb{Z}_{0}^{n}}k_{n+1}^{\frac{n}{2}}\left[\sum _{l=1}^n (k_lL_l)^2\right] ^{-\frac{n}{4}}\\
&\times K_{\frac{n}{2}}\left(\frac{2\pi k_{n+1}}{\beta}\sqrt{\sum
_{l=1}^n(k_lL_l)^2}\right) +\frac{1-n}{2\beta}E_{n}^{\prime}(0;\frac{1}{L_1},\ldots ,\frac{1}{L_n}) \\
&-\frac{1}{\beta}\sum _{\substack{ j,k=1\\j<k }}^n \ln{\Theta _1 \left( 0|\frac{\sqrt{-1}\beta ^{-1}L_j^2L_k^2}{2\pi V} \right)
}-\frac{1}{\beta}\sum _{j=1}^{n}\ln{\Theta _1 \left( \langle\theta _j\rangle |\frac{2\pi\sqrt{-1}\beta ^{-1}V}{L_j^2}\right)},
\end{split}\label{free-energy-tot-high}
\end{equation}
with the three constants $\kappa_1=\frac{n^2-3n+4}{4}$, $\kappa_2=\frac{n^2-7n+8}{4}$ and $\kappa_3=\frac{n^2-7n+4}{4}$. These constants are
non vanishing for all dimensions $n$.\par

We want to emphasize that the expressions for both $\mathcal{F}_{low}^{\theta}$ and $\mathcal{F}_{high}^{\theta}$ are valid for all $\beta$ and
$V$, that is $\mathcal{F}_{high}^{\theta}=\mathcal{F}_{low}^{\theta}$. However, their difference lies in the different polynomial structure as
function of the temperature.\par

Evidently, \eqref{constraint} is satisfied so that for the uniform $n$-torus the sum of the last two terms in \eqref{free-energy-tot-high},
denoted by $f_{high}^{\theta}(\beta ;V)$, admits the following series expansion

\begin{multline}
 f_{high}^{\theta}(\beta ;V)=-\frac{n(n-1)}{2\beta}\left(\sum_{m=1}^{\infty}\frac{1}{m}\ \frac{1}{1-e^{m\beta
^{-1}V^{\frac{4-n}{n}}}}+2\sum_{m=1}^{\infty}\frac{(-1)^m}{m}\ \frac{e^{\frac{1}{2}m\beta ^{-1}V^{\frac{4-n}{n}}}}{1-e^{m\beta
^{-1}V^{\frac{4-n}{n}}}}\right)\\ -\frac{1}{\beta}\left(\sum_{m=1}^{\infty}\frac{1}{m}\ \frac{1}{1-e^{(2\pi)^2m\beta ^{-1}V^{\frac{n-2}{n}}}}
+2\sum_{j=1}^{n}\sum_{m=1}^{\infty}\frac{(-1)^m}{m}\ \frac{e^{\frac{(2\pi)^2}{2}m\beta ^{-1}V^{\frac{n-2}{n}}}\cos (2\pi m\langle\theta
_j\rangle)}{1-e^{(2\pi)^2m\beta ^{-1}V^{\frac{n-2}{n}}}}\right).
\end{multline}
All terms show an exponential decrease for $\beta\rightarrow 0$. The same is true for the various derivatives of $f_{high}^{\theta}$ which
appear in the different thermodynamic functions. Since $K_{\nu}(z)\simeq\sqrt{\frac{\pi}{2z}}e^{-z}\left(1+\mathcal{O}(\frac{1}{z})\right)$ for
$|z|\rightarrow\infty$ \cite{gradshteyn}, the summands in the fourth term of \eqref{free-energy-tot-high} show an exponential decrease as
$\beta\rightarrow 0$.\par

As has been stated, \eqref{free-energy-tot-high} is exact. But what would be the result, if the general formula \eqref{K3} was used instead of
\eqref{free-energy-tot-high} for determing the high-temperature limit? Taking into account \eqref{zeta-torus} and the following explicit
expressions for the Seeley coefficients of the Laplace operator

\begin{equation}
a_m(\Delta _1^{\mathbb{T}^{n}}|_{im d_{2}^{\ast}})= \begin{cases} (4\pi)^{-\frac{n}{2}}\ (n-1)\ V &\text{if $m=0$}\\ 1-n &\text{if $m=n$}\\
0 &\text{if $m\neq 0, m\neq n$},\end{cases}\label{seeley-torus}
\end{equation}
we would have obtained \eqref{free-energy-tot-high} up to the exponentially decreasing fourth term.\par

In the high temperature regime the internal energy is given by

\begin{equation}
\begin{split}
U_{high}^{\theta}(\beta ;V)= & n(n-1) \pi^{-\frac{n+1}{2}}\ \Gamma (\frac{n+1}{2})\ \zeta _{R}(n+1)
\beta ^{-(n+1)}\ V+\frac{\kappa_1}{\beta}\\
&-4\pi (n-1)\beta ^{-\frac{n+4}{2}}V^{\frac{n+2}{2n}}\sum _{k_{n+1}=1}^{\infty}\ \sum _{\vec{k}\in\mathbb{Z}_{0}^{n}}k_{n+1}^{\frac{n+2}{2}}\
|\vec{k}|^{\frac{2-n}{2}}K_{\frac{n}{2}-1}\left(\frac{2\pi V^{\frac{1}{n}}}{\beta}k_{n+1}|\vec{k}|\right) \\
&+\frac{\partial}{\partial\beta}\left(\beta f_{high}^{\theta}(\beta ;V)\right),
\end{split}\label{energy-tot-high-V}
\end{equation}
where the relation $\frac{d}{d z}K_{\nu}(z)=-K_{\nu -1}(z)-\frac{\nu}{z}K_{\nu}(z)$ have been used (see \cite{gradshteyn}). Apart from the
leading Stefan-Boltzmann term all other terms in \eqref{energy-tot-high-V} show a exponential decrease in the high temperature limit. In
addition there exists a term linear in the temperature which is positive for all dimensions $n$. A lengthy calculation finally gives the
following expression for the entropy

\begin{equation}
\begin{split}
S_{high}^{\theta}(\beta ;V) &=(n^2-1) \pi^{-\frac{n+1}{2}}\ \Gamma (\frac{n+1}{2})\ \zeta _{R}(n+1)\beta ^{-n}\ V+\kappa_1(1-\ln{\beta})+\\
&+2(n-1)\beta ^{-\frac{n}{2}}V^{\frac{1}{2}}\sum _{k_{n+1}=1}^{\infty}\ \sum
_{\vec{k}\in\mathbb{Z}_{0}^{n}}k_{n+1}^{\frac{n}{2}}|\vec{k}|^{-\frac{n}{2}}K_{\frac{n}{2}}\left(\frac{2\pi
V^{\frac{1}{n}}}{\beta}k_{n+1}|\vec{k}|\right) \\ &-4\pi (n-1)\beta ^{-\frac{n+2}{2}}V^{\frac{n+2}{2n}}\sum _{k_{n+1}=1}^{\infty}\ \sum
_{\vec{k}\in\mathbb{Z}_{0}^{n}}k_{n+1}^{\frac{n+2}{2}}|\vec{k}|^{\frac{2-n}{2}}K_{\frac{n}{2}-1}\left(\frac{2\pi
V^{\frac{1}{n}}}{\beta}k_{n+1}|\vec{k}|\right) \\ &+\frac{n-1}{2}E_{n}^{\prime}(0;1,\ldots ,1)-\kappa_3\ln{2\pi}-\kappa_4\ln{V}
\\&+\beta^{2}\ \frac{\partial}{\partial\beta}f_{high}^{\theta}(\beta ;V).
\end{split}\label{entropy-tot-high-V}
\end{equation}
where $\kappa_4=\kappa_2+\frac{n-1}{n}$. The pressure in the high temperature regime finally reads

\begin{equation}
\begin{split}
P_{high}^{\theta}(\beta ;V) &=(n-1) \pi^{-\frac{n+1}{2}}\ \Gamma (\frac{n+1}{2})\ \zeta _{R}(n+1)\beta ^{-(n+1)}-\frac{\kappa_4}{\beta V}
\\
&-4\pi\frac{(n-1)}{n} \beta ^{-\frac{n+4}{2}}V^{\frac{2-n}{2n}}\sum _{k_{n+1}=1}^{\infty}\ \sum
_{\vec{k}\in\mathbb{Z}_{0}^{n}}k_{n+1}^{\frac{n+2}{2}}\ |\vec{k}|^{\frac{2-n}{2}}K_{\frac{n}{2}-1}\left(\frac{2\pi
V^{\frac{1}{n}}}{\beta}k_{n+1}|\vec{k}|\right)\\
&-\frac{\partial}{\partial V}f_{high}^{\theta}(\beta ;V).
\end{split}\label{pressure-tot-high-V}
\end{equation}
The term linear in temperature and volume is caused by the harmonic part of the gauge field on $\mathbb{T}^n$. It vanishes for $n=2$ and is
positive for $n=3,4$.

\section{Summary}

To summarize, we want to highlight the main results of this paper: We shed some light onto the relationship between the topology of the
compact, closed and connected spatial manifold $X$ and the thermodynamic properties of quantum Maxwell theory at finite temperature. This has
been achieved by providing a rigorous determination of the free energy for this system. It has been shown that there are inequivalent quantum
theories giving rise to $\theta$-vacua, which are related to the first Betti number of $X$. The vacuum energy is the sum of the energy of the
transverse modes and the energy of the harmonic modes of the Maxwell field. The finite temperature excitations are affected by the topology of
$X$ unless both the first and second cohomology of $X$ vanish.

In the Hamilton approach the quantization has been performed directly on the physical phase space using an appropriate parametrization of that
space. The minimal subtraction scheme was applied to regularize the vacuum energy of the transverse modes.\par

The functional integral quantization method on the other hand requires some advanced steps in order to overcome the Gribov problem: We
constructed an integrable and globally defined measure on the space of all thermal gauge fields and introduced a consistent reduction procedure
to eliminate the gauge degrees of freedom. By expanding the zeta function of the Laplace operators on $\mathbb{T}_{\beta}^1\times X$ in terms
of the zeta function associated to the corresponding Laplace operators on the spatial manifold $X$ we succeeded in proving the equivalence with
the Hamiltonian approach.\par

The thermodynamics of a photon gas confined to a $n$-torus is explicitly elaborated as a special example. Exact expressions for the
thermodynamic functions for both the low- and high temperature regimes are derived and the validity of the Nernst theorem is proved.\par

In light of the ongoing interest in the thermodynamic structure of quantum fields on spaces with non-trivial topologies, it will be a subject
of future work to extend the present analysis to manifolds with boundaries and to apply these results for studying further the topological
aspects of the Casimir effect in space-time models with compactified extra dimensions.

\section{Acknowledgements}
I would like to express my gratitude to H. H\"{u}ffel for his encouragement and his valuable comments.
\bigskip\bigskip

\appendix
\section{Appendix: Riemann Theta functions} In this section we want to summarize the main facts regarding the Riemann Theta function. Let us now introduce the
$r$-dimensional Riemann Theta function with characteristics $a,b\in\mathbb{R}^{r}$ by

\begin{equation}
\Theta _r\begin{bmatrix} a \\b\end{bmatrix} (u|B)=\sum\limits _{m\in\mathbb{Z}^r}\exp{\lbrace \pi\sqrt{-1}(m+a) ^{\dag}\cdot B\cdot
(m+a)+2\pi\sqrt{-1}(m+a)^{\dag}\cdot (u+b)\rbrace },\label{theta-original}
\end{equation}
for a symmetric complex $r\times r$ dimensional square matrix $B$ whose imaginary part is positive definite, $u\in\mathbb{C}^r$ and the
superscript $\dag$ denotes the transpose. It has the modular property

\begin{equation}
\Theta _r\begin{bmatrix} a+k \\b+l\end{bmatrix} (u|B)=e^{2\pi\sqrt{-1}a^{\dag}l}\ \Theta _r\begin{bmatrix} a\\b\end{bmatrix} (u|B),\qquad
k,l\in\mathbb{Z}^r.\label{modular-property}
\end{equation}
We write $\Theta _{r}(u|B):=\Theta _r\begin{bmatrix} 0 \\0\end{bmatrix} (u|B)$ for the Riemann Theta function with characteristics $(0,0)$. It
can be easily shown that $\Theta _r(-u|B)=\Theta _r(u|B)$ which leads to $\Theta _r\begin{bmatrix} -u\\0\end{bmatrix} (0|B)=\Theta
_r\begin{bmatrix} u\\0\end{bmatrix} (0|B)$. By using the Poisson sum formula one can derive the following important "duality" formula

\begin{equation}
\Theta _r (u|B)=\det{(-\sqrt{-1}B)}^{-\frac{1}{2}}\ \Theta _r\begin{bmatrix} u \\0\end{bmatrix} (0|-B^{-1}).\label{duality}
\end{equation}
In order to provide a suitable asymptotic expansion of the Riemann Theta function we recall the existence of the infinite product
representation (Jacobi triple product) which reads

\begin{equation}
\Theta _1(u|B)=\prod_{k=1}^{\infty}(1-q^{2k})(1+q^{2k-1}e^{2\pi\sqrt{-1}u})(1+q^{2k-1}e^{-2\pi\sqrt{-1}u}),\label{jacobi}
\end{equation}
where $q=e^{\pi\sqrt{-1}B}$. Let us now take $B=\sqrt{-1}\tau$, with $\tau\in\mathbb{R}$ and $\tau >0$. If $\tau$ and $u=\Re u+\sqrt{-1}\Im u$
satisfy the constraint

\begin{equation}
\tau - 2|\Im u|\geq 0,\label{constraint}
\end{equation}
then one can apply the series expansion of the logarithm and the geometric series formula to \eqref{jacobi}. A straightforward calculation
finally gives

\begin{equation}
\begin{split}
\ln{\Theta _1(u|\sqrt{-1}\tau)}
&=-\sum_{k=1}^{\infty}\sum_{m=1}^{\infty}\frac{q^{2km}}{m}+\sum_{k=1}^{\infty}\sum_{m=1}^{\infty}\frac{(-1)^{m+1}}{m}q^{(2k-1)m}\left(
e^{2\pi\sqrt{-1}mu} +e^{-2\pi\sqrt{-1}mu}\right) \\ &= \sum_{m=1}^{\infty}\frac{1}{m}\ \frac{1}{1-e^{2\pi\tau
m}}+\sum_{m=1}^{\infty}\frac{(-1)^{m}}{m}\ \frac{e^{\pi\tau m}\left( e^{2\pi\sqrt{-1}mu}+e^{-2\pi\sqrt{-1}mu}\right)}{1-e^{2\pi\tau
m}}.\label{asymptotic-expansion}
\end{split}
\end{equation}

\section{Appendix: Epstein zeta functions}
In this appendix the main results on Epstein zeta functions \cite{Epstein1, Epstein2} (e.g. see also \cite{Lim-Teo, elizalde}) are summarized.
Let us introduce the homogeneous Epstein zeta function $E_{m}$ in the variables $(c_1,\ldots ,c_m)\in\mathbb{R}^{m}$ by

\begin{equation}
E_{m}(s;c_1,\ldots ,c_m)=\sum _{(k_1,\ldots ,k_m)\in\mathbb{Z}_0^m}\left[ \sum _{i=1}^m (c_ik_i)^2\right]^{-s},\qquad
s\in\mathbb{C},\label{epstein-zeta}
\end{equation}
where $\mathbb{Z}_0^m:=\mathbb{Z}^m\backslash\{0\}$. The sum is convergent for $\Re(s)>\frac{m}{2}$. The analytic continuation of the Epstein
zeta function is regular on the whole complex $s$-plane up to a unique pole at $s=\frac{m}{2}$. Under a constant scale transformation
$c_i\mapsto \lambda c_i$, $\lambda\in\mathbb{R}$, the Epstein function transforms as

\begin{equation}
E_{m}(s;\lambda c_1,\ldots ,\lambda c_m)=\lambda ^{-2s}E_{m}(s;c_1,\ldots ,c_m).\label{epstein-trafo}
\end{equation}
Epstein zeta functions in different dimensions are related by the Chowla-Selberg formula,

\begin{equation}
\begin{split}
E_{m}(s;c_1,\ldots ,c_m)= &E_{l}(s;c_1,\ldots ,c_l)+\frac{\pi ^{\frac{l}{2}}\Gamma (s-\frac{l}{2})}{\prod _{i=1}^{l}c_i\Gamma
(s)}E_{m-l}(s-\frac{l}{2};c_{l+1},\ldots ,c_m)\\ &+\frac{1}{\Gamma (s)}T_{m,l}(s;c_1,\ldots ,c_m),
\end{split}
\end{equation}
where
\begin{equation}
\begin{split}
T_{m,l}(s;c_1,\ldots ,c_m) = &\frac{2\pi ^{s}}{\prod _{i=1}^{l}c_i}\ \sum _{(k_1,\ldots ,k_l)\in\mathbb{Z}_0^l}\ \sum _{(k_{l+1},\ldots
,k_m))\in\mathbb{Z}_0^{m-l}}\left[ \frac{\sum_{i=1}^{l}(\frac{k_i}{c_i})^2}{\sum_{i=l+1}^{m}(k_ic_i)^2} \right]^{\frac{2s-l}{4}}\\
&\times K_{s-\frac{l}{2}}\left(2\pi\sqrt{\left(\sum_{i=1}^{l}(\frac{k_i}{c_i})^2\right)\left(\sum_{i=l+1}^{m}(k_ic_i)^2\right)}\right).
\end{split}
\end{equation}
Here $K_{\nu}(z)$ denotes the modified Bessel function of the second kind \cite{gradshteyn}. The function $s\mapsto T_{m,l}(s;c_1,\ldots ,c_m)$
is analytic on $\mathbb{C}$. Using the Chowla-Selberg formula one can prove the so-called reflection formula

\begin{equation}
\pi ^{-s}\Gamma (s)E_{m}(s;c_1,\ldots ,c_m)=\frac{\pi ^{s-\frac{m}{2}}}{\prod _{i=1}^{m}c_i}\Gamma
(\frac{m}{2}-s)E_{m}(\frac{m}{2}-s;\frac{1}{c_1},\ldots ,\frac{1}{c_m}),\label{reflection}
\end{equation}
from which the following results follows

\begin{equation}
E_{m}(0;c_1,\ldots ,c_m)=E_{1}(0;c_1)=2\zeta _{R}(0)=-1.\label{epstein=0}
\end{equation}
It follows from \eqref{epstein-trafo} and \eqref{epstein=0} that the derivative of $E_m$ transforms under constant scale transformations
according to

\begin{equation}
E_{m}^{\prime}(0;\lambda c_1,\ldots ,\lambda c_m)=2\ln{\lambda}+E_{m}^{\prime}(0;c_1,\ldots ,c_m),
\end{equation}
where $\zeta _{R}(s)=\sum _{k=1}^nk^{-s}$ denotes the Riemann zeta function. By recursion we can express the Epstein zeta function in terms of
the Riemann zeta function as follows:

\begin{equation}
\begin{split}
E_{m}(s;c_1,\ldots ,c_m) &=2c_1^{-2s}\zeta _{R}(2s)+\frac{2}{\Gamma (s)}\sum_{i=1}^{m-1}\frac{\pi^{\frac{i}{2}}\Gamma
(s-\frac{i}{2})}{c_{i+1}^{2s-i}\prod _{j=1}^{i}c_j}\zeta _{R}(2s-i) \\ &+\frac{4\pi^{s}}{\Gamma (s)}\sum_{i=1}^{m-1}\frac{1}{\prod
_{j=1}^{i}c_j}\sum_{(k_1,\ldots
,k_i)\in\mathbb{Z}_0^{i}}\sum_{k_{i+1}=1}^{\infty}\frac{\left[\sum_{j=1}^{i}(\frac{k_j}{c_j})^2\right]^{\frac{s}{2}-\frac{i}{4}}}{\left[
k_{i+1}c_{i+1} \right]^{s-\frac{i}{2}}} \\ &\times K_{s-\frac{i}{2}}\left( 2\pi k_{i+1}c_{i+1} \sqrt{\sum_{j=1}^{i}(\frac{k_j}{c_j})^2}\right).
\end{split}
\end{equation}
Since $\lim _{s\rightarrow 0}(\frac{d}{ds}\frac{f(s)}{\Gamma (s)})=0$ for a well-behaved function $f(s)$ and using the reflection formula one
obtains for the derivative of the Epstein zeta function $E_{m}^{\prime}(0;c_1,\ldots ,c_m):=\frac{\partial}{\partial
s}|_{s=0}E_{m}(s;c_1,\ldots ,c_m)$ in $s=0$

\begin{equation}
E_{m}^{\prime}(0;c_1,\ldots ,c_m)=E_{l}^{\prime}(0;c_1,\ldots ,c_l)+\frac{\pi ^{-\frac{m}{2}}\Gamma (\frac{m}{2})}{\prod
_{i=1}^{m}c_i}E_{m-l}(\frac{m}{2};\frac{1}{c_{l+1}},\ldots ,\frac{1}{c_m})+T_{m,l}(0;c_1,\ldots ,c_m).\label{epstein-expansion}
\end{equation}

\end{document}